\documentclass[9pt]{IEEEtran}

\usepackage{cite}
\ifCLASSINFOpdf
\else
\fi

\usepackage{amsmath}
\usepackage{amssymb}
\usepackage{amsfonts}
\usepackage{bm}

\usepackage{algorithm}
\usepackage{algpseudocode}

\usepackage{array}

\usepackage[colorlinks]{hyperref}
\hypersetup{
    colorlinks=true,
    linkcolor=black,
    filecolor=magenta,
    citecolor=black,      
    }

\newtheorem{assumption}{Assumption}
\newtheorem{remark}{Remark}

\newtheorem{theorem}{Theorem}
\newtheorem{corollary}{Corollary}
\newtheorem{proposition}{Proposition}

\newtheorem{claim}{Claim}

\usepackage{graphicx}
\usepackage{float}
\usepackage{xcolor}
\usepackage{enumitem}

\begin{document}

\title{Set-Membership Filter for Discrete-Time Nonlinear Systems Using State Dependent Coefficient Parameterization}

\author{Diganta~Bhattacharjee and~Kamesh~Subbarao, \IEEEmembership{Senior Member, IEEE}
\thanks{D. Bhattacharjee and K. Subbarao are with the Department of Mechanical and Aerospace Engineering, The University of Texas at Arlington, Arlington, TX, 76019 USA e-mail: diganta.bhattacharjee@mavs.uta.edu, subbarao@uta.edu}}
\maketitle

\begin{abstract}
In this technical note, a recursive set-membership filtering algorithm for discrete-time nonlinear dynamical systems subject to unknown but bounded process and measurement noises is proposed. The nonlinear dynamics is represented in a pseudo-linear form using the state dependent coefficient (SDC) parameterization. Matrix Taylor expansions are utilized to expand the state dependent matrices about the state estimates. Upper bounds on the norms of remainders in the matrix Taylor expansions are calculated on-line using a non-adaptive random search algorithm at each time step. Utilizing these upper bounds and the ellipsoidal set description of the uncertainties, a two-step filter is derived that utilizes the `correction-prediction' structure of the standard Kalman Filter variants. At each time step, correction and prediction ellipsoids are constructed that contain the true state of the system by solving the corresponding semi-definite programs (SDPs). Finally, a simulation example is included to illustrate the effectiveness of the proposed approach.   
\end{abstract}


%

\section{Introduction}
A broad class of state estimation and filtering approaches require stochastic description of the process noise corrupting the state of the system and the noise associated with the measurements, e.g., Kalman Filters are optimal for linear systems with uncorrelated, Gaussian process and measurement noises \cite{Anderson_Moore_1979}. 
An alternative approach involves assuming the noises to be unknown but bounded. Under this assumption, 
the solution to the estimation problem requires 
set of states consistent with the knowledge of the bounds on the noises, the governing equations or models, and the measurements \cite{Polyak_et_al_2004, Becis-Aubry_et_al_2008}. This approach is generally referred to as set-membership, set-valued, guaranteed estimation or filtering \cite{Maksarov_Norton_1996, El_Ghaoui_Calafiore_2001,  Polyak_et_al_2004, Becis-Aubry_et_al_2008, Yang_Li_IEEE_Trans_Auto_Control_2009, Shamma_Tu_1997, Calafiore_2005} and was introduced in the late 1960s and early 1970s \cite{Witsenhausen_1968, Schweppe_1968, Bertsekas_Rhodes_1971}. As the actual set containing all possible values of the state is, in general, very complex and hard to obtain, several approximations (ellipsoids, polytopes, zonotopes) have been studied in the literature. 


In this technical note, the ellipsoidal state estimation problem is considered and the terminology set-membership filter (SMF) is adopted. Over the years, set-membership filtering for linear systems has attracted significant attention and the theory is well-established (see, e.g., \cite{El_Ghaoui_Calafiore_2001, Becis-Aubry_et_al_2008, Yang_Li_IEEE_Trans_Auto_Control_2009} and the references therein). Particularly, the filter design proposed in this technical note is motivated by \cite{El_Ghaoui_Calafiore_2001, Yang_Li_IEEE_Trans_Auto_Control_2009} where the set estimation problems were converted into recursive algorithms that require solutions to semi-definite programs (SDPs) at each time step. Recently, several extensions of this approach have emerged in the literature (see, e.g.,\cite{Wei_et_al_2015, Wang_et_al_2018, Wang_et_al_2019}). 

On the other hand, set-membership filtering for discrete-time nonlinear systems has received less attention. For discrete-time nonliner systems, similar to the Extended Kalman Filter (EKF), set-membership filtering approaches typically involve linearizing the nonlinear dynamics about the state estimate trajectory \cite{Scholte_Campbell_2003, Zhou_et_al_2008, Calafiore_2005, Wang_et_al_2018}. An extended set-membership filter (ESMF) was developed in \cite{Scholte_Campbell_2003} by linearizing the state dynamics about the state estimates and bounding the linearization errors using interval analysis. An improvement over the algorithm proposed in \cite{Scholte_Campbell_2003} was provided in \cite{Zhou_et_al_2008}. The SDP-based approach for discrete-time nonlinear systems was introduced in \cite{Calafiore_2005} with a prediction-correction form. Recently, this approach was extended in \cite{Wang_et_al_2018} where the linearization errors were bounded by ellipsoids on-line.
 
Alternatively, state dependent coefficient (SDC) parameterization can be utilized to represent a nonlinear system in \textit{a pseudo-linear form} with state dependent system matrices \cite{Mracek_Cloutier_1998, Cimen2012survey}. The parameterization is non-unique and the non-uniqueness can be utilized to enhance performance of the controller or filter design (see, e.g., \cite{Cimen2012survey, Dani_et_al_2014}). SDC parameterization has been utilized for filter design in a stochastic framework for discrete-time nonlinear systems (see \cite{Jaganath_et_al_2005, Chang_2018}). However, set-membership filtering using the SDC parameterization has not been addressed in the existing open literature to the best of the authors' knowledge.

Motivated by the above discussion, a recursive SMF utilizing the SDC parameterization (SDC-SMF) is proposed in this technical note for discrete-time nonlinear systems subject to unknown but bounded process and measurement noises. A two-step correction-prediction form is developed, similar to the Kalman Filter variants \cite{Anderson_Moore_1979}. The proposed filter requires solutions to two SDPs at each time step, similar to \cite{Calafiore_2005, Wang_et_al_2018}. The technical novelties of the proposed approach are three fold and are summarized as follows.
\begin{enumerate}
\item A single SDC parameterization of the nonlinear system is utilized to obtain a pseudo-linear representation which preserves the nonlinearity in the governing equations. To the best of our knowledge, this is the first SMF for discrete-time nonlinear systems that utilizes the SDC parameterization.
\item 
Instead of taking the EKF-like approach for SMF frameworks as in \cite{Scholte_Campbell_2003, Calafiore_2005, Zhou_et_al_2008, Wang_et_al_2018}, the state dependent matrices are expanded about the state estimates in matrix Taylor expansions using Vetter calculus \cite{Vetter_1973}. Upper bounds on the norms of remainders in the matrix Taylor expansions are calculated \textit{on-line} at each time step and those bounds are utilized in the filter design at every recursion. This approach is different from the approaches in \cite{Scholte_Campbell_2003, Zhou_et_al_2008} where interval analysis were utilized to bound the linearization errors and from the recent approach in \cite{Wang_et_al_2018} where the linearization errors were bounded by ellipsoidal sets.
\item We make use of predefined structures for the state estimates which are similar to the traditional Kalman filtering technique. However, this is unlike the SDP-based SMFs for discrete-time nonlinear systems in \cite{Calafiore_2005, Wang_et_al_2018} where the state estimates constitute a part of the optimization variables. 
\end{enumerate}
Our approach of dealing with the remainders is computationally cheaper compared to the ellipsoidal bounding methodology in \cite{Wang_et_al_2018} which requires additional SDPs to be solved, increasing the associated computational cost. Also, the above technical novelties lead to better filtering performance compared to the framework in \cite{Wang_et_al_2018}, as demonstrated in the simulation results.
The rest of this technical note is organized as follows. Section \ref{Preliminaries and Problem Formulation} describes the preliminaries and problem formulation for the SDC-SMF. Section \ref{Main Results} discusses the main results for the proposed SDC-SMF and formulates the SDPs to be solved at each time step to find the ellipsoidal sets containing the true state of the system. Finally, Section \ref{Simulation Example} includes a simulation example and Section \ref{Conclusion} presents the concluding remarks. 
\subsubsection*{Notation}
The symbol $\mathbb{Z}_{\star}$ denotes the set of non-negative integers. For a square matrix $\bm{X}$, the notation $\bm{X} > 0$ (respectively, $\bm{X} \geq 0$) means $\bm{X}$ is symmetric and positive definite (respectively, positive semi-definite). Similarly, $\bm{X} < 0$ (respectively, $\bm{X} \leq 0$) means $\bm{X}$ is symmetric and negative definite (respectively, negative semi-definite). The notations $\text{diag}(\cdot)$, $\bm{I}_n$, $\bm{O}_n$, and $\bm{0}_{n}$ denote block-diagonal matrices, the $n \times n$ identity matrix, the $n \times n$ null matrix, and the vector of zeros of dimension $n$, respectively. The symbol $||\cdot||$ denotes the spectral norm for matrices and the Euclidean norm for vectors. Ellipsoids are denoted by $\mathcal{E}(\bm{c}, \bm{P}) = \{ \bm{x} \in \mathbb{R}^n :  (\bm{x} - \bm{c})^{\text{T}} \bm{P}^{-1} (\bm{x} - \bm{c}) \leq 1 \}$ where $\bm{c} \in \mathbb{R}^n$ is the center of the ellipsoid and $\bm{P} > 0$ is the \textit{shape matrix} that characterizes the orientation and size of the ellipsoid in $\mathbb{R}^n$. Also, the Kronecker product is denoted by $\otimes$ and a function that is continuously differentiable $t$ times is said to be of class $C^t$. To this end, $C^0$ denotes the class of continuous functions.
\section{Preliminaries and Problem Formulation}  \label{Preliminaries and Problem Formulation}
Consider discrete-time, nonlinear dynamical systems of the form
\begin{equation} \label{general nonlinear dynamics}
\begin{split}
\bm{x}_{k+1} &= \bm{f}_d (\bm{x}_k) + \bm{w}_k \\ 
\bm{y}_k &= \bm{h}_d (\bm{x}_k) + \bm{v}_k 
\end{split}
\end{equation}
where $k \in \mathbb{Z}_{\star}$, $\bm{x}_k \in \mathbb{R}^n$ is the state of the system, $\bm{w}_k \in \mathbb{R}^n$ is the process noise or input disturbance, $\bm{y}_k \in \mathbb{R}^p$ is the measured output, and $\bm{v}_k \in \mathbb{R}^p$ is the measurement noise. We make the following standing assumption on the nonlinear functions $\bm{f}_d : \mathbb{R}^n \rightarrow \mathbb{R}^n$ and $\bm{h}_d : \mathbb{R}^n \rightarrow \mathbb{R}^p$.
\begin{assumption} \label{Assumptions on the nonlinear functions}
$\bm{f}_d (\bm{0}_n) = \bm{0}_n$, $\bm{h}_d (\bm{0}_n) = \bm{0}_p$, and $\bm{f}_d (\cdot) \in C^t$, $\bm{h}_d (\cdot) \in C^t$ where $t \geq 2$.
\end{assumption}
Under Assumption \ref{Assumptions on the nonlinear functions}, the nonlinear functions can be put into corresponding pseudo-linear forms using the SDC parameterization as
\begin{equation} \label{SDC form calculation-1}
\begin{split}
\bm{f}_d(\bm{x}_k) &= \bm{A}(\bm{x}_k) \bm{x}_k \\ 
\bm{h}_d(\bm{x}_k) &= \bm{H}(\bm{x}_k) \bm{x}_k
\end{split}
\end{equation}
where $\bm{A}:\mathbb{R}^n \rightarrow \mathbb{R}^{n \times n}$ and $\bm{H}:\mathbb{R}^n \rightarrow \mathbb{R}^{p \times n}$ are nonlinear matrix-valued functions. To this end, we recall the following useful result.
\begin{proposition} \cite{Cimen2012survey, Vidyasagar_2002} \label{Proposition regarding the SDC parameterization}
Under Assumption \ref{Assumptions on the nonlinear functions}, SDC parameterizations of $\bm{f}_d(\bm{x}_k), \ \bm{h}_d(\bm{x}_k)$ as in \eqref{SDC form calculation-1} always exist for some $C^{t-1}$ matrix-valued functions $\bm{A}:\mathbb{R}^n \rightarrow \mathbb{R}^{n \times n}$ and $\bm{H}:\mathbb{R}^n \rightarrow \mathbb{R}^{p \times n}$. This property is satisfied by the following parameterizations
\begin{equation} \label{SDC form calculation-2}
\begin{split}
\bm{A}(\bm{x}_k) &= \int_{0}^{1} \frac{\partial \bm{f}_d(\bm{x}_k)}{\partial \bm{x}_k} \Bigg\rvert_{\bm{x}_k = \lambda \bm{x}_k} d \lambda \\
\bm{H}(\bm{x}_k) &= \int_{0}^{1} \frac{\partial \bm{h}_d(\bm{x}_k)}{\partial \bm{x}_k} \Bigg\rvert_{\bm{x}_k = \lambda \bm{x}_k} d \lambda
\end{split}
\end{equation}
where $\lambda$ is a dummy variable of integration. The parameterizations in \eqref{SDC form calculation-2} are guaranteed to exist under Assumption \ref{Assumptions on the nonlinear functions}. Furthermore, any SDC parameterization of $\bm{f}_d(\bm{x}_k), \ \bm{h}_d(\bm{x}_k)$ as in \eqref{SDC form calculation-1} satisfies $\bm{A}(\bm{0}_n) = \frac{\partial \bm{f}_d(\bm{x}_k)}{\partial \bm{x}_k} \big\rvert_{\bm{x}_k = \bm{0}_n} $, $\bm{H}(\bm{0}_n) = \frac{\partial \bm{h}_d(\bm{x}_k)}{\partial \bm{x}_k} \big\rvert_{\bm{x}_k = \bm{0}_n}$.
\end{proposition}
Note that multiple SDC parameterizations of the form \eqref{SDC form calculation-1} are possible for $n>1$ using mathematical factorization\cite{Cimen2012survey}. However, we choose the SDC parameterizations given in \eqref{SDC form calculation-2} under Assumption \ref{Assumptions on the nonlinear functions} and describe the nonlinear system \eqref{general nonlinear dynamics} in an equivalent pseudo-linear form as 
\begin{equation} \label{SDC form of the nonlinear dynamics}
\begin{split}
\bm{x}_{k+1} &= \bm{A}(\bm{x}_k) \bm{x}_k + \bm{w}_k \\ 
\bm{y}_k &= \bm{H}(\bm{x}_k) \bm{x}_k + \bm{v}_k .
\end{split}
\end{equation}  
For a detailed discussion on the SDC parameterization, refer to \cite{Cimen2012survey, Mracek_Cloutier_1998} and references therein. We make the following assumption on the state dynamics of system \eqref{SDC form of the nonlinear dynamics}.
\begin{assumption}   \label{State dynamics compactness assumption}
\cite[Section V]{Shamma_Tu_1997} There exist compact sets $\mathbb{D}_0, \mathbb{D} \subset \mathbb{R}^n$ and $\epsilon_1 > 0$ such that $\bm{x}_0 \in \mathbb{D}_0$ implies
\begin{displaymath}
\bm{x}_k + \epsilon_1 \mathcal{B}(\bm{x}_k) \subset \mathbb{D}, \quad \forall k \in \mathbb{Z}_{\star} 
\end{displaymath}
where $\mathcal{B}(\bm{x}_k)$ is the closed unit ball in $\mathbb{R}^n$ centered at $\bm{x}_k$.
\end{assumption}
The above assumption implies that the state $\bm{x}_k$ evolves within a compact set $\mathbb{D}$ which is not necessarily small \cite{Shamma_Tu_1997}. Now, we state the following assumptions for system \eqref{SDC form of the nonlinear dynamics} where $\mathbb{D}_0$ is as described in Assumption \ref{State dynamics compactness assumption}.
\begin{assumption} \label{Assumption: initial conditions, noises, and trajectory for the true system}
\begin{enumerate}[label= 3.\arabic*]
\item \label{Assumption 3 - initial conditions} $\bm{x}_0$ is unknown but belongs to a known ellipsoid, i.e., $\bm{x}_0 \in \mathcal{E} (\hat{\bm{x}}_0, \bm{P}_0) \subseteq \mathbb{D}_0$ where $\hat{\bm{x}}_0$ is a given initial estimate and $\bm{P}_0$ is known.
\item \label{Assumption 3 - process and measurement noise ellipsoids} $\bm{w}_k$ and $\bm{v}_k$ are unknown but bounded and belong to known ellipsoids, i.e., $\bm{w}_k \in \mathcal{E}(\bm{0}_n, \bm{Q}_k)$ and $\bm{v}_k \in \mathcal{E}(\bm{0}_p, \bm{R}_k), \ \forall k \in \mathbb{Z}_{\star}$ where $\bm{Q}_k$, $\bm{R}_k$ are known and satisfy $||\bm{Q}_k|| \leq q$ and $||\bm{R}_k|| \leq r, \forall k \in \mathbb{Z}_{\star}$ with some $q, r > 0$.
\end{enumerate}
\end{assumption}
Assumption \ref{Assumption 3 - process and measurement noise ellipsoids} means that the process and measurement noises in system \eqref{SDC form of the nonlinear dynamics} are uniformly upper bounded. Now, we introduce the final assumption on system \eqref{SDC form of the nonlinear dynamics}.
\begin{assumption}  \label{Observability related assumption}
Along any trajectory of system \eqref{SDC form of the nonlinear dynamics} under Assumption \ref{State dynamics compactness assumption}, define
\begin{equation}
\begin{split}
\bm{\phi}_{k+s,k} &= \bm{A}(\bm{x}_{k+s-1}) \ \bm{A}(\bm{x}_{k+s-2}) \cdots \bm{A}(\bm{x}_{k}) \\
\mathcal{O}_{k,k+s} &= 
\begin{bmatrix}
\bm{H}(\bm{x}_k) \\
\bm{H}(\bm{x}_{k+1}) \bm{\phi}_{k+1,k} \\
\vdots \\
\bm{H}(\bm{x}_{k+s}) \bm{\phi}_{k+s,k}
\end{bmatrix}.
\end{split}
\end{equation} 
for any $s \in \mathbb{Z}_\star \backslash\{0\}$. There exists an $N_o \in \mathbb{Z}_\star \backslash\{0\}$ such that
\begin{equation} \label{Observability rank condition}
\text{rank} \left( \mathcal{O}_{k,k+N_o-1} \right) = n, \quad \forall k \in \mathbb{Z}_\star.
\end{equation}
\end{assumption}
This is an observability assumption where $N_o = n$ might be possible. The above assumption leads to the following result.
\begin{proposition}
Under Assumption \ref{Observability related assumption}, there exist $\mu_1, \mu_2 > 0$ such that
\begin{equation}
\mu_1 \bm{I}_n \leq \mathcal{O}_{k,k+N_o-1}^\text{T} \mathcal{O}_{k,k+N_o-1} \leq \mu_2 \bm{I}_n.
\end{equation}
\end{proposition}
\begin{IEEEproof}
Follows directly from Proposition 5.1 in \cite{Shamma_Tu_1997} (or see the proof of Proposition 4.1 in \cite[Section 4]{Song_Grizzle_1995}).
\end{IEEEproof} 
\begin{remark}
With the knowledge of set $\mathbb{D}$, Assumption \ref{Observability related assumption} requires one to check if the rank condition in \eqref{Observability rank condition} is satisfied for all $\bm{x}_k \in \mathbb{D}$ with some $N_o \in \mathbb{Z}_\star \backslash\{0\}$. This can be done by carrying out a theoretical analysis (cf., Section \ref{Simulation Example}) or by implementing a numerical routine.
\end{remark}
\begin{remark} \label{Remark on theoretical properties of the SDC-SMF}
With our compactness and observability assumptions, theoretical properties of the proposed SDC-SMF for system \eqref{SDC form of the nonlinear dynamics} (in a sense similar to that of Definition 3.1 in \cite{Shamma_Tu_1997}) can be assessed by appropriately modifying the analysis and results in \cite{Shamma_Tu_1997} (cf., Section \ref{Theoretical properties of SDC-SMF}).
\end{remark}
\subsection{SDC-SMF Objectives}
The objective is to develop an SDC-SMF for system \eqref{SDC form of the nonlinear dynamics} having a correction-prediction form, similar to the Kalman Filter variants \cite{Anderson_Moore_1979}. This helps to obtain an accurate estimate of the state and a reliable evaluation of the estimation error. The filtering objectives are as follows. 
\subsubsection{Correction Step} 
At each time step $k \in \mathbb{Z}_{\star}$, upon receiving the measurement $\bm{y}_{k}$ with $\bm{v}_{k} \in \mathcal{E} (\bm{0}_p, \bm{R}_{k})$ and given $\bm{x}_{k} \in \mathcal{E} (\hat{\bm{x}}_{k|k-1}, \bm{P}_{k|k-1})$, the objective is to find a \textit{correction ellipsoid} such that $\bm{x}_{k} \in \mathcal{E} (\hat{\bm{x}}_{k|k}, \bm{P}_{k|k})$. The corrected state estimate is given by 
\begin{equation} \label{objective-corrected state estimate}
\hat{\bm{x}}_{k|k} = \hat{\bm{x}}_{k|k-1} + \bm{L}_{k} (\bm{y}_{k} - \bm{H}(\hat{\bm{x}}_{k|k-1}) \hat{\bm{x}}_{k|k-1})
\end{equation}
where $\bm{L}_k$ is the filter gain.
\subsubsection{Prediction Step}
At each time step $k \in \mathbb{Z}_{\star}$, given $\bm{x}_{k} \in \mathcal{E} (\hat{\bm{x}}_{k|k}, \bm{P}_{k|k})$ and $\bm{w}_k \in \mathcal{E} (\bm{0}_n, \bm{Q}_k)$, the objective is to find a \textit{prediction ellipsoid} such that $\bm{x}_{k+1} \in \mathcal{E} (\hat{\bm{x}}_{k+1|k}, \bm{P}_{k+1|k})$ where the predicted state estimate is given by 
\begin{equation} \label{objective-predicted state estimate}
\hat{\bm{x}}_{k+1|k} = \bm{A} (\hat{\bm{x}}_{k|k}) \hat{\bm{x}}_{k|k} . 
\end{equation}
Initialization is provided by $\hat{\bm{x}}_{0|-1} = \hat{\bm{x}}_0$ and $\bm{P}_{0|-1} = \bm{P}_0$ \cite{Anderson_Moore_1979} which form the \textit{initial prediction ellipsoid} due to Assumption \ref{Assumption 3 - initial conditions}. 
\subsection{Matrix Taylor Expansions of the SDC Matrices}
Assume that the state of system \eqref{SDC form of the nonlinear dynamics} at time step $k$ belongs to the prediction ellipsoid of time step $k-1$, i.e., $\bm{x}_k \in \mathcal{E} (\hat{\bm{x}}_{k|k-1}, \bm{P}_{k|k-1})$ where $\hat{\bm{x}}_{k|k-1}$ and $\bm{P}_{k|k-1}$ are known. Then, there exists a $\bm{z}_{k|k-1} \in \mathbb{R}^n$ with $||\bm{z}_{k|k-1}||\leq 1$ such that
\begin{equation} \label{True state}
\bm{x}_k = \hat{\bm{x}}_{k|k-1} + \bm{E}_{k|k-1} \bm{z}_{k|k-1}
\end{equation}
where $\bm{E}_{k|k-1}$ is the Cholesky factorization of $\bm{P}_{k|k-1}$, i.e., $\bm{P}_{k|k-1} = \bm{E}_{k|k-1} \bm{E}_{k|k-1}^\textnormal{T}$ \cite{El_Ghaoui_Calafiore_2001, Yang_Li_IEEE_Trans_Auto_Control_2009}. Utilizing the matrix Taylor expansion in \cite{Vetter_1973}, $\bm{H}({\bm{x}}_k) = \bm{H}(\hat{\bm{x}}_{k|k-1} + \bm{E}_{k|k-1} \bm{z}_{k|k-1})$ can be expanded about the state estimate $\hat{\bm{x}}_{k|k-1}$ as 
\begin{equation} \label{Expansion of H(x)}
\begin{split}
\bm{H}({\bm{x}}_k) =& \hspace{0.1cm} \bm{H}(\hat{\bm{x}}_{k|k-1}) + \bm{K}_1 (\hat{\bm{x}}_{k|k-1}) \bm{\Delta}_1 (\bm{\xi}_{k|k-1}) \\
                    & + \bm{R}_{\bm{H}_{2}} (\hat{\bm{x}}_{k|k-1}, \bm{x}_k)
\end{split}
\end{equation}
where $\bm{K}_1 (\hat{\bm{x}}_{k|k-1}) = \mathcal{D}_{\bm{x}^\text{T}} \bm{H}(\hat{\bm{x}}_{k|k-1})$ is the derivative matrix evaluated at $\hat{\bm{x}}_{k|k-1}$, $\bm{\Delta}_1 (\bm{\xi}_{k|k-1}) = \left(\bm{\xi}_{k|k-1} \otimes \bm{I}_n \right)$ with $\bm{\xi}_{k|k-1} = \bm{x}_k - \hat{\bm{x}}_{k|k-1} = \bm{E}_{k|k-1} \bm{z}_{k|k-1}$, and $\bm{R}_{\bm{H}_{2}} (\hat{\bm{x}}_{k|k-1}, \bm{x}_{k})$ is the remainder (see Section 6 in \cite{Vetter_1973}). Similarly, the matrix $\bm{A}({\bm{x}}_k) = \bm{A}(\hat{\bm{x}}_{k|k} + \bm{E}_{k|k} \bm{z}_{k|k})$ is expanded as 
\begin{equation} \label{Expansion of A(x)}
\begin{split}
\bm{A}(\bm{x}_{k}) =& \ \bm{A}(\hat{\bm{x}}_{k|k}) + \bm{K}_2 (\hat{\bm{x}}_{k|k}) \bm{\Delta}_2 (\bm{\xi}_{k|k}) \\
                    & + \bm{R}_{\bm{A}_{2}} (\hat{\bm{x}}_{k|k}, \bm{x}_{k})
\end{split}
\end{equation} 
where $\bm{K}_2 (\hat{\bm{x}}_{k|k}) = \mathcal{D}_{\bm{x}^\text{T}} \bm{A}(\hat{\bm{x}}_{k|k})$, $\bm{\Delta}_2 (\bm{\xi}_{k|k}) = \left( \bm{\xi}_{k|k} \otimes \bm{I}_n \right)$ with $\bm{\xi}_{k|k} = \bm{x}_k - \hat{\bm{x}}_{k|k} = \bm{E}_{k|k} \bm{z}_{k|k}$ where $\bm{P}_{k|k} = \bm{E}_{k|k} \bm{E}_{k|k}^\textnormal{T}$ and $||\bm{z}_{k|k}|| \leq 1$. As $\bm{A}(\bm{x}_k), \ \bm{H}(\bm{x}_k)$ are calculated using \eqref{SDC form calculation-2} under Assumption \ref{Assumptions on the nonlinear functions}, $\bm{K}_1 (\cdot), \ \bm{K}_2 (\cdot)$ are at least continuous matrix-valued functions.
\subsection{Upper Bounds on the Norms of Remainders in Matrix Taylor Expansions}
At each time step, upper bounds on the norms of the remainders in \eqref{Expansion of H(x)}-\eqref{Expansion of A(x)} are calculated and utilized in the SDC-SMF design. Thus, we require the following quantities:
\begin{equation}
\begin{split}
\bar{r}_{A_{k}} &= \sup_{\bm{x}_{k} \in \mathcal{E} (\hat{\bm{x}}_{k|k}, \bm{P}_{k|k})} || \bm{R}_{\bm{A}_{2}} (\hat{\bm{x}}_{k|k}, \bm{x}_k)|| \\
\bar{r}_{H_{k}} &= \sup_{\bm{x}_{k} \in \mathcal{E} (\hat{\bm{x}}_{k|k-1}, \bm{P}_{k|k-1})} || \bm{R}_{\bm{H}_{2}} (\hat{\bm{x}}_{k|k-1}, \bm{x}_{k}) || .
\end{split}
\end{equation}
Next, we state an important result regarding $\bar{r}_{A_{k}}$ and $\bar{r}_{H_{k}}$.
\begin{proposition} \label{Proposition: boundedness of the remainders}
$\bar{r}_{A_{k}}$ and $\bar{r}_{H_{k}}$ belong to the boundaries of the ellipsoids $\mathcal{E} (\hat{\bm{x}}_{k|k}, \bm{P}_{k|k})$ and $\mathcal{E} (\hat{\bm{x}}_{k|k-1}, \bm{P}_{k|k-1})$, respectively. 
\end{proposition} 
\begin{IEEEproof}
Let us denote $\mathcal{E}_{c_{k}} = \mathcal{E} (\hat{\bm{x}}_{k|k}, \bm{P}_{k|k})$ and $\mathcal{E}_{p_{k}} = \mathcal{E} (\hat{\bm{x}}_{k|k-1}, \bm{P}_{k|k-1})$. Due to the continuity of the matrix-valued functions and compactness of the ellipsoids, we have
\begin{displaymath}
\begin{split}
h_k &= \sup_{\bm{x} \in \mathcal{E}_{p_{k}}} ||\bm{H}({\bm{x}})||, \quad k_{1_{k}} = \sup_{\bm{x} \in \mathcal{E}_{p_{k}}} ||\bm{K}_1 ({\bm{x}})||, \\
a_k &= \sup_{\bm{x} \in \mathcal{E}_{c_{k}}} ||\bm{A}({\bm{x}})||, \hspace{0.4cm} k_{2_{k}} = \sup_{\bm{x} \in \mathcal{E}_{c_{k}}} ||\bm{K}_2 ({\bm{x}})||
\end{split} 
\end{displaymath}
where $0 < a_k, h_k, k_{1_{k}}, k_{2_{k}} < \infty$. Now, consider the remainder $\bm{R}_{\bm{A}_{2}} (\hat{\bm{x}}_{k|k}, \bm{x}_k)$ in \eqref{Expansion of A(x)}, expressed as
\begin{equation}   \label{R_A expression for the remainder norm max calculation}
\bm{R}_{\bm{A}_{2}} (\hat{\bm{x}}_{k|k}, \bm{x}_k ) = \bm{A}(\bm{x}_k) - \bm{A}(\hat{\bm{x}}_{k|k}) - \bm{K}_2 \bm{\Delta}_2 
\end{equation}
where the arguments of $\bm{K}_2 (\cdot)$ and $\bm{\Delta}_2 (\cdot)$ have been dropped. Taking the norm leads to
\begin{displaymath}
\begin{split}
|| \bm{R}_{\bm{A}_{2}} (\hat{\bm{x}}_{k|k}, \bm{x}_k)|| \leq 2 a_k + k_{2_{k}} ||\bm{\Delta}_2 ||. 
\end{split}
\end{displaymath}
Utilizing $\bm{\Delta}_2 = \left( \bm{\xi}_{k|k} \otimes \bm{I}_n \right)$, the following holds:
\begin{equation} \label{Delta 2 constraint}
\begin{split}
||\bm{\Delta}_2|| = ||\bm{\xi}_{k|k}|| \ ||\bm{I}_n|| \leq ||\bm{E}_{k|k} || \ || \bm{z}_{k|k}||.
\end{split}                                                                                                              
\end{equation}
Denoting $||\bm{E}_{k|k}|| = \gamma_{k|k}$, \eqref{Delta 2 constraint} becomes $||\bm{\Delta}_2|| \leq \gamma_{k|k} || \bm{z}_{k|k}||$. Then, the norm of the remainder satisfies
\begin{equation} \label{Remainder of A: upper bound}
\begin{split}
|| \bm{R}_{\bm{A}_{2}} (\hat{\bm{x}}_{k|k}, \bm{x}_k)|| \leq & 2 a_k + k_{2_{k}} \gamma_{k|k} || \bm{z}_{k|k}||. 
\end{split}																		     
\end{equation}
Clearly, the upper bound corresponds to $|| \bm{z}_{k|k}|| = 1$, i.e., $\bar{r}_{A_{k}}$ belongs to the boundary of $\mathcal{E}_{c_{k}}$. Carrying out a similar analysis for $\bm{R}_{\bm{H}_{2}} (\hat{\bm{x}}_{k|k-1}, \bm{x}_{k})$ yields 
\begin{displaymath}
|| \bm{R}_{\bm{H}_{2}} (\hat{\bm{x}}_{k|k-1}, \bm{x}_{k}) || \leq 2 h_k + k_{1_{k}} \gamma_{k|k-1} ||\bm{z}_{k|k-1}||
\end{displaymath}
with $||\bm{E}_{k|k-1}|| = \gamma_{k|k-1}$ which shows that  $\bar{r}_{H_{k}}$ belongs to the boundary of $\mathcal{E}_{p_{k}}$. This completes the proof.
\end{IEEEproof}
Therefore, $\bar{r}_{A_{k}}$ can be obtained by solving the optimization problem
\begin{equation}  \label{Norm maximization optimization}
\begin{split}
& \sup_{\bm{z}_{k|k}} || \bm{R}_{\bm{A}_{2}} (\hat{\bm{x}}_{k|k}, \hat{\bm{x}}_{k|k} + \bm{E}_{k|k} \bm{z}_{k|k})|| \\
& \text{subject to} \ ||\bm{z}_{k|k}|| = 1
\end{split}
\end{equation}
where the feasible set is non-convex. The non-convex problem can be convexified and solved using the primal-dual methods numerically (see, e.g., \cite{Bertsekas_1979}). Alternatively, a much simpler approach, so-called \textit{non-adaptive random search algorithm} \cite{Tempo_et_al_2005, Brooks_1958}, can be utilized to obtain an approximate solution to \eqref{Norm maximization optimization}. Adopting this approach, the norm of the remainder is evaluated $N$ times by randomly sampling $N$ number of points on the unit circle $||\bm{z}_{k|k}|| = 1$. Then, the upper bound on the remainder norm is given by the \textit{empirical maximum} \cite{Tempo_et_al_2005} as
\begin{equation}  \label{r_A}
\begin{split}
r_{A_{k}} =  \max_{i = 1,2,...,N} || \bm{R}_{\bm{A}_{2}} (\hat{\bm{x}}_{k|k}, \hat{\bm{x}}_{k|k} + \bm{E}_{k|k} \bm{z}_{{k|k}_{i}})||
\end{split}
\end{equation}
where $||\bm{z}_{{k|k}_{i}}|| = 1, \ i = 1,2,...,N$. Similarly, the upper bound on the norm of $\bm{R}_{\bm{H}_{2}} (\hat{\bm{x}}_{k|k-1}, \bm{x}_{k})$ is determined as
\begin{equation}  \label{r_H}
\begin{split}
r_{H_{k}} = \max_{i = 1,2,...,N} || \bm{R}_{\bm{H}_{2}} (\hat{\bm{x}}_{k|k-1}, \hat{\bm{x}}_{k|k-1} + \bm{E}_{k|k-1} \bm{z}_{k|{k-1}_{i}})||
\end{split}
\end{equation}
where $||\bm{z}_{k|{k-1}_{i}}|| = 1, \ i=1,2,...,N$. Moreover, as $N \rightarrow \infty$, we have $r_{A_{k}} \rightarrow \bar{r}_{A_{k}}$ and $r_{H_{k}} \rightarrow \bar{r}_{H_{k}}$ (see Theorem 7.4 in \cite{Tempo_et_al_2005}). 
The arguments of $\bm{K}_i (\cdot)$ and $\bm{\Delta}_i (\cdot) \ (i=1,2)$ have been dropped in the subsequent analysis.

\begin{remark}  \label{Remark on the form of the system on which the filter design is based}
Using the matrix Taylor expansions in \eqref{Expansion of H(x)}-\eqref{Expansion of A(x)}, the governing equations utilized for the SDC-SMF design for system \eqref{SDC form of the nonlinear dynamics} can be expressed as
\begin{equation} \label{dynamics utilized for filter design}
\begin{split}
\bm{x}_{k+1} &= \bm{A}(\hat{\bm{x}}_{k|k}) \bm{x}_k + \tilde{\bm{w}}_k \\
\bm{y}_{k} &= \bm{H}(\hat{\bm{x}}_{k|k-1}) \bm{x}_{k} + \tilde{\bm{v}}_{k} \\
\end{split}
\end{equation}
where 
\begin{displaymath} 
\begin{split}
\tilde{\bm{w}}_k &= \bm{w}_k + \bm{K}_2 \bm{\Delta}_2 \bm{x}_k + \bm{R}_{\bm{A}_{2}} (\hat{\bm{x}}_{k|k}, \bm{x}_k) \bm{x}_k \\
\tilde{\bm{v}}_{k} &= \bm{v}_{k} + \bm{K}_1 \bm{\Delta}_1 \bm{x}_k +  \bm{R}_{\bm{H}_{2}} (\hat{\bm{x}}_{k|k-1}, \bm{x}_{k}) \bm{x}_k .
\end{split}
\end{displaymath}
The governing equations in \eqref{dynamics utilized for filter design} are different from the governing equations utilized in EKF-like approach-based SMF frameworks (see, e.g.,  Section 3 in \cite{Scholte_Campbell_2003}).
The bounds on the terms in $\tilde{\bm{w}}_k$, $\tilde{\bm{v}}_k$ and the ellipsoidal set description of the true state $\bm{x}_k$ are utilized in the next section to derive the SDC-SMF.
\end{remark}
\section{Main Results}  \label{Main Results}
This section formulates the SDPs to be solved at each time step for the correction and prediction steps. The arguments of $\bm{R}_{\bm{A}_{2}} (\cdot)$ and $\bm{R}_{\bm{H}_{2}} (\cdot)$ are omitted in the subsequent analysis for notational simplicity. With that, let us state Theorem \ref{Theorem:correction step} that summarizes the filtering problem at the correction step.
\begin{theorem} \label{Theorem:correction step}
Consider system \eqref{SDC form of the nonlinear dynamics} under Assumptions \ref{Assumption 3 - initial conditions} and \ref{Assumption 3 - process and measurement noise ellipsoids}. At each time step $k \in \mathbb{Z}_{\star}$, upon receiving the measurement $\bm{y}_{k}$ with $\bm{v}_{k} \in \mathcal{E} (\bm{0}_p, \bm{R}_{k})$ and given $\bm{x}_{k} \in \mathcal{E} (\hat{\bm{x}}_{k|k-1}, \bm{P}_{k|k-1})$, the state $\bm{x}_{k}$ is contained in the optimal correction ellipsoid $\mathcal{E} (\hat{\bm{x}}_{k|k}, \bm{P}_{k|k})$, if there exist $\bm{P}_{k|k} > 0$, $\bm{L}_{k}$, $\tau_i \geq 0, \ i=1, 2, 3, 4, 5, 6$ as solutions to the following SDP:
\begin{equation} \label{The complete problem statement-1}
\begin{split} 
& \min_{\bm{P}_{k|k}, \bm{L}_{k}, \tau_{1},\tau_{2},\tau_{3}, \tau_{4}, \tau_{5}, \tau_{6}} \hspace{0.2cm} \text{trace}(\bm{P}_{k|k}) \\
& \text{subject to} \\
& \bm{P}_{k|k} > 0 \\
& \tau_i \geq 0, \ i = 1, 2, 3, 4, 5, 6 \\
& \begin{bmatrix}
-\bm{P}_{k|k} & \bm{\Pi}_{k|k-1} \\ \\
\bm{\Pi}^T_{k|k-1} & -\bm{\Theta} (\tau_{1},\tau_{2},\tau_{3}, \tau_{4}, \tau_{5}, \tau_{6})
\end{bmatrix} \leq 0
\end{split}
\end{equation} 
where $\bm{\Pi}_{k|k-1}$ and $\bm{\Theta} (\tau_{1},\tau_{2},\tau_{3}, \tau_{4}, \tau_{5}, \tau_{6})$ are given by
\begin{equation} \label{Pi_k_k-1 and Theta(tau_1,...tau_6)}
\begin{split}  
\bm{\Pi}&_{k|k-1} \\
     =& \ \Big [\bm{0}_{n} \quad (\bm{E}_{k|k-1} - \bm{L}_{k} \bm{H} (\hat{\bm{x}}_{k|k-1})\bm{E}_{k|k-1}) \quad - \bm{L}_{k} \\
      & - \bm{L}_{k} \bm{K}_1 \quad - \bm{L}_{k} \quad - \bm{L}_{k} \bm{K}_1 \quad - \bm{L}_{k} \Big] \\ 
\bm{\Theta} & (\tau_1, \tau_{2},\tau_{3}, \tau_{4}, \tau_{5},\tau_{6}) \\
 =& \ \textnormal{diag} \ (1- \tau_1 - \tau_{2} - \tau_{5} \gamma_{k|k-1}^2 \hat{\bm{x}}_{k|k-1}^\textnormal{T} \hat{\bm{x}}_{k|k-1} \\
& - \tau_{6} r_{H_{k}}^{{2}} \hat{\bm{x}}_{k|k-1}^\textnormal{T} \hat{\bm{x}}_{k|k-1}, \tau_{1} \bm{I}_n \\
& - \tau_{3} \gamma_{k|k-1}^2 \bm{E}_{k|k-1}^\textnormal{T} \bm{E}_{k|k-1} - \tau_{4} r_{H_{k}}^{{2}} \bm{E}_{k|k-1}^\textnormal{T} \bm{E}_{k|k-1}, \\
& \tau_{2} \bm{R}_{k}^{-1},  \tau_{3} \bm{I}_{n^2}, \tau_{4} \bm{I}_{p}, \tau_{5} \bm{I}_{n^2}, \tau_{6} \bm{I}_{p}).
\end{split}
\end{equation}
Furthermore, center of the correction ellipsoid is given by the corrected state estimate in \eqref{objective-corrected state estimate}.
\end{theorem}
\begin{IEEEproof}
Utilizing the corrected state estimate in \eqref{objective-corrected state estimate}, the estimation error at the correction step is
\begin{equation} \label{estimation error at the correction step}
\begin{split}
\bm{x}_{k} & - \hat{\bm{x}}_{k|k} \\
           =& (\bm{x}_{k} - \hat{\bm{x}}_{k|k-1}) -  \bm{L}_{k} (\bm{y}_{k} - \bm{H}(\hat{\bm{x}}_{k|k-1}) \hat{\bm{x}}_{k|k-1})  \\
				  =& \bm{E}_{k|k-1} \bm{z}_{k|k-1} - \bm{L}_{k} \Big[(\bm{H}(\hat{\bm{x}}_{k|k-1}) + \bm{K}_1 \bm{\Delta}_1 + \bm{R}_{\bm{H}_{2}}) \\
					 &  \times (\hat{\bm{x}}_{k|k-1} + \bm{E}_{k|k-1} \bm{z}_{k|k-1}) - \bm{H}(\hat{\bm{x}}_{k|k-1})  \hat{\bm{x}}_{k|k-1} + \bm{v}_{k} \Big] \\
					 =& \left( \bm{E}_{k|k-1} - \bm{L}_{k} \bm{H} (\hat{\bm{x}}_{k|k-1})\bm{E}_{k|k-1} \right) \bm{z}_{k|k-1} - \bm{L}_{k} \bm{v}_{k}  \\
					  & - \bm{L}_{k} \bm{K}_1 \bm{\Delta}_1 \bm{E}_{k|k-1} \bm{z}_{k|k-1} - \bm{L}_{k} \bm{R}_{\bm{H}_{2}} \bm{E}_{k|k-1} \bm{z}_{k|k-1}  \\
					  & - \bm{L}_{k} \bm{K}_1 \bm{\Delta}_1 \hat{\bm{x}}_{k|k-1} - \bm{L}_{k} \bm{R}_{\bm{H}_{2}} \hat{\bm{x}}_{k|k-1}
\end{split}
\end{equation}
Denote the unknowns in \eqref{estimation error at the correction step} as 
\begin{equation}
\begin{split}
\bm{\Delta}_{3} &= \bm{\Delta}_1 \bm{E}_{k|k-1} \bm{z}_{k|k-1} \in \mathbb{R}^{n^2}  \\
\bm{\Delta}_{4} &= \bm{R}_{\bm{H}_{2}} \bm{E}_{k|k-1} \bm{z}_{k|k-1} \in \mathbb{R}^{p}  \\
\bm{\Delta}_{5} &= \bm{\Delta}_1 \hat{\bm{x}}_{k|k-1}  \in \mathbb{R}^{n^2} \\
\bm{\Delta}_{6} &= \bm{R}_{\bm{H}_{2}} \hat{\bm{x}}_{k|k-1} \in \mathbb{R}^{p}. \\
\end{split}
\end{equation}
Next, define a vector of all the unknowns in \eqref{estimation error at the correction step} as 
\begin{equation}
\bm{\zeta} = \Big[
1 \quad \bm{z}_{k|k-1}^{\text{T}} \quad \bm{v}_{k}^{\text{T}} \quad \bm{\Delta}_{3}^{\text{T}} \quad \bm{\Delta}_{4}^{\text{T}} \quad \bm{\Delta}_{5}^{\text{T}} \quad \bm{\Delta}_{6}^{\text{T}} \Big ]^{\text{T}}.
\end{equation}
Therefore, the estimation error in \eqref{estimation error at the correction step} can be expressed in terms of $\bm{\zeta}$ as
\begin{equation}
\bm{x}_{k} - \hat{\bm{x}}_{k|k-1} = \bm{\Pi}_{k|k-1} \bm{\zeta}
\end{equation}
where $\bm{\Pi}_{k|k-1}$ is as shown in \eqref{Pi_k_k-1 and Theta(tau_1,...tau_6)}. Now, $\bm{x}_{k} \in \mathcal{E} (\hat{\bm{x}}_{k|k}, \bm{P}_{k|k})$ can be expressed as 
\begin{equation} \label{State estimation error constraint}
\begin{split}
\bm{\zeta}^{\text{T}}  & \Big[ \bm{\Pi}_{k|k-1}^{\text{T}} \bm{P}_{k|k}^{-1} \bm{\Pi}_{k|k-1} \\
& - \text{diag} (1, \bm{O}_n, \bm{O}_p, \bm{O}_{n^2}, \bm{O}_{p}, \bm{O}_{n^2}, \bm{O}_p) \Big] \bm{\zeta} \leq 0.
\end{split} 
\end{equation}
Using the definition of $\bm{\Delta}_1$, it can be shown that $||\bm{\Delta}_1|| \leq \gamma_{k|k-1}$ (Similar to \eqref{Delta 2 constraint}). With that, the following inequalities hold:
\begin{displaymath}
\begin{split}
\begin{cases}
\bm{\Delta}_3^\text{T} \bm{\Delta}_3 = & \bm{z}_{k|k-1}^{\text{T}} \bm{E}_{k|k-1}^{\text{T}} \bm{\Delta}_1^\text{T} \bm{\Delta}_1 \bm{E}_{k|k-1} \bm{z}_{k|k-1} \\
                                       & \leq \gamma_{k|k-1}^2 \bm{z}_{k|k-1}^{\text{T}} \bm{E}_{k|k-1}^{\text{T}} \bm{E}_{k|k-1} \bm{z}_{k|k-1} , \\
\bm{\Delta}_5^\text{T} \bm{\Delta}_5 =& \hat{\bm{x}}_{k|k-1}^\text{T} \bm{\Delta}_1^\text{T} \bm{\Delta}_1 \hat{\bm{x}}_{k|k-1} \leq \gamma_{k|k-1}^2 \hat{\bm{x}}_{k|k-1}^\text{T} \hat{\bm{x}}_{k|k-1} . 
\end{cases}                                      
\end{split}
\end{displaymath}
Similarly, utilizing the upper bound on the norm of remainder $\bm{R}_{\bm{H}_{2}}$, the following inequalities are derived:
\begin{displaymath}
\begin{split}
\begin{cases}
 \bm{\Delta}_4^\text{T} \bm{\Delta}_4 =& \bm{z}_{k|k-1}^{\text{T}} \bm{E}_{k|k-1}^{\text{T}} \bm{R}_{\bm{H}_{2}}^\text{T} \bm{R}_{\bm{H}_{2}} \bm{E}_{k|k-1} \bm{z}_{k|k-1} \\
                                       & \leq r_{H_{k}}^{2} \bm{z}_{k|k-1}^{\text{T}} \bm{E}_{k|k-1}^{\text{T}} \bm{E}_{k|k-1} \bm{z}_{k|k-1}, \\
 \bm{\Delta}_6^\text{T} \bm{\Delta}_6 =& \hat{\bm{x}}_{k|k-1}^\text{T} \bm{R}_{\bm{H}_{2}}^\text{T} \bm{R}_{\bm{H}_{2}} \hat{\bm{x}}_{k|k-1}^\text{T} \leq r_{H_{k}}^{2} \hat{\bm{x}}_{k|k-1}^\text{T} \hat{\bm{x}}_{k|k-1} .
\end{cases}
\end{split}
\end{displaymath}
Therefore, all the unknowns in $\bm{\zeta}$ should satisfy the following inequalities 
\begin{equation*}
\begin{split}
\begin{cases}
\bm{z}_{k|k-1}^{\text{T}} \bm{z}_{k|k-1} -1 \leq 0, \\
\bm{v}_{k}^T \bm{R}_k^{-1} \bm{v}_{k}  -1 \leq 0 ,\\
\bm{\Delta}_{3}^{\text{T}} \bm{\Delta}_{3} - \gamma_{k|k-1}^2 \bm{z}_{k|k-1}^{\text{T}} \bm{E}_{k|k-1}^{\text{T}} \bm{E}_{k|k-1} \bm{z}_{k|k-1} \leq 0 ,\\
\bm{\Delta}_{4}^{\text{T}} \bm{\Delta}_{4} - r_{H_{k}}^2 \bm{z}_{k|k-1}^{\text{T}} \bm{E}_{k|k-1}^{\text{T}} \bm{E}_{k|k-1} \bm{z}_{k|k-1}  \leq 0,\\
\bm{\Delta}_{5}^{\text{T}} \bm{\Delta}_{5} - \gamma_{k|k-1}^2 \hat{\bm{x}}_{k|k-1}^{\text{T}} \hat{\bm{x}}_{k|k-1} \leq 0 ,\\
\bm{\Delta}_{6}^{\text{T}} \bm{\Delta}_{6} - r_{H_{k}}^2  \hat{\bm{x}}_{k|k-1}^{\text{T}} \hat{\bm{x}}_{k|k-1}  \leq 0. 
\end{cases}
\end{split}
\end{equation*}
The above inequalities are expressed in terms of $\bm{\zeta}$ as follows
\begin{equation}   \label{Constraints on the unknown variables}
\begin{split}
\begin{cases}
\bm{\zeta}^{\text{T}} \text{diag} (-1, \bm{I}_n, \bm{O}_p, \bm{O}_{n^2}, \bm{O}_{p}, \bm{O}_{n^2}, \bm{O}_p) \bm{\zeta} \leq 0 ,\\
\bm{\zeta}^{\text{T}} \text{diag} (-1, \bm{O}_n, \bm{R}_{k}^{-1}, \bm{O}_{n^2}, \bm{O}_{p}, \bm{O}_{n^2}, \bm{O}_p) \bm{\zeta} \leq 0 ,\\
\bm{\zeta}^{\text{T}} \text{diag} (0, -\gamma_{k|k-1}^2 \bm{E}_{k|k-1}^{\text{T}} \bm{E}_{k|k-1}, \bm{O}_p, \bm{I}_{n^2}, \bm{O}_{p}, \\ 
\hspace{1cm} \bm{O}_{n^2}, \bm{O}_p) \bm{\zeta} \leq 0 ,\\
\bm{\zeta}^{\text{T}} \text{diag} (0, -r_{H_{k}}^2 \bm{E}_{k|k-1}^{\text{T}} \bm{E}_{k|k-1}, \bm{O}_p, \bm{O}_{n^2}, \bm{I}_{p}, \\
\hspace{1cm} \bm{O}_{n^2}, \bm{O}_p) \bm{\zeta} \leq 0 ,\\
\bm{\zeta}^{\text{T}} \text{diag} (-\gamma_{k|k-1}^2 \hat{\bm{x}}_{k|k-1}^{\text{T}} \hat{\bm{x}}_{k|k-1}, \bm{O}_n, \bm{O}_p, \bm{O}_{n^2},  \\
\hspace{1cm} \bm{O}_{p}, \bm{I}_{n^2}, \bm{O}_p) \bm{\zeta} \leq 0 ,\\
\bm{\zeta}^{\text{T}} \text{diag} (- r_{H_{k}}^2  \hat{\bm{x}}_{k|k-1}^{\text{T}} \hat{\bm{x}}_{k|k-1}, \bm{O}_n, \bm{O}_p, \bm{O}_{n^2}, \bm{O}_{p}, \\
\hspace{1cm} \bm{O}_{n^2}, \bm{I}_p) \bm{\zeta} \leq 0 .\\
\end{cases}
\end{split}
\end{equation}
Next, the S-procedure (see, e.g., \cite{Boyd_et_al_1994}) is applied to the inequalities in \eqref{State estimation error constraint} and \eqref{Constraints on the unknown variables}. The inequality in \eqref{State estimation error constraint} holds if there exist $\tau_1 \geq 0, \tau_2 \geq 0, \tau_3 \geq 0, \tau_4 \geq 0, \tau_5 \geq 0, \tau_6 \geq 0$ such that the following is true :
\begin{displaymath}
\begin{split}
& \bm{\Pi}_{k|k-1}^{\text{T}} \bm{P}_{k|k}^{-1} \bm{\Pi}_{k|k-1} - \text{diag} (1, \bm{O}_n, \bm{O}_p, \bm{O}_{n^2}, \bm{O}_{p}, \bm{O}_{n^2}, \bm{O}_p) \\
& - \tau_1 \text{diag} (-1, \bm{I}_n, \bm{O}_p, \bm{O}_{n^2}, \bm{O}_{p}, \bm{O}_{n^2}, \bm{O}_p) \\
& - \tau_2 \text{diag} (-1, \bm{O}_n, \bm{R}_{k}^{-1}, \bm{O}_{n^2}, \bm{O}_{p}, \bm{O}_{n^2}, \bm{O}_p) \\
& - \tau_3 \text{diag} (0, -\gamma_{k|k-1}^2 \bm{E}_{k|k-1}^{\text{T}} \bm{E}_{k|k-1}, \bm{O}_p, \bm{I}_{n^2}, \bm{O}_{p}, \bm{O}_{n^2}, \bm{O}_p) \\
& -\tau_4 \text{diag} (0, -r_{H_{k}}^2 \bm{E}_{k|k-1}^{\text{T}} \bm{E}_{k|k-1}, \bm{O}_p, \bm{O}_{n^2}, \bm{I}_{p}, \bm{O}_{n^2}, \bm{O}_p) \\
& -\tau_5 \text{diag} (-\gamma_{k|k-1}^2 \hat{\bm{x}}_{k|k-1}^{\text{T}} \hat{\bm{x}}_{k|k-1}, \bm{O}_n, \bm{O}_p, \bm{O}_{n^2}, \bm{O}_{p}, \bm{I}_{n^2}, \bm{O}_p) \\
& - \tau_6 \text{diag} (- r_{H_{k}}^2  \hat{\bm{x}}_{k|k-1}^{\text{T}} \hat{\bm{x}}_{k|k-1}, \bm{O}_n, \bm{O}_p, \bm{O}_{n^2}, \bm{O}_{p}, \bm{O}_{n^2}, \bm{I}_p) \\
& \leq 0.
\end{split}
\end{displaymath}
The above inequality can be expressed in a compact form as
\begin{equation} \label{Inequality-compact form}
\begin{split}
\bm{\Pi}_{k|k-1}^{\text{T}} \bm{P}_{k|k}^{-1} \bm{\Pi}_{k|k-1} - \bm{\Theta} (\tau_1,\tau_2,\tau_3, \tau_4, \tau_5, \tau_6) \leq 0
\end{split}
\end{equation}
where $\bm{\Theta} (\tau_1,\tau_2,\tau_3, \tau_4, \tau_5, \tau_6)$ is as given in \eqref{Pi_k_k-1 and Theta(tau_1,...tau_6)}. Utilizing the Schur complement (see, e.g., \cite{Boyd_et_al_1994}), the inequality in \eqref{Inequality-compact form} can be equivalently expressed as
\begin{equation}   \label{Sufficient condition-prediction}
\begin{bmatrix}
-\bm{P}_{k|k} & & \bm{\Pi}_{k|k-1} \\ \\
\bm{\Pi}^T_{k|k-1} & & -\bm{\Theta} (\tau_1,\tau_2,\tau_3, \tau_4, \tau_5, \tau_6)
\end{bmatrix} \leq 0.
\end{equation}
Solving the inequality in \eqref{Sufficient condition-prediction} with $\bm{P}_{k|k} > 0$ and $\tau_i \geq 0, \ i = 1, 2, 3, 4, 5, 6$ yields \textit{a correction ellipsoid} that contains the true state of the system. To obtain the minimal set containing the true state, the sum of the squared lengths of semi-axes of the correction ellipsoid is minimized by minimizing the trace of $\bm{P}_{k|k}$. This completes the proof.
\end{IEEEproof}

The next Theorem summarizes the filtering problem at the prediction step. 

\begin{theorem}  \label{Theorem:prediction step}
Consider system \eqref{SDC form of the nonlinear dynamics} under Assumption \ref{Assumption 3 - process and measurement noise ellipsoids} with the current state 
$\bm{x}_k \in \mathcal{E} (\hat{\bm{x}}_{k|k}, \bm{P}_{k|k})$ and $\bm{w}_k \in \mathcal{E} (\bm{0}_n, \bm{Q}_k)$. Then, the successor state $\bm{x}_{k+1}$ belongs to the optimal prediction ellipsoid $\mathcal{E} (\hat{\bm{x}}_{k+1|k}, \bm{P}_{k+1|k})$, if there exist $\bm{P}_{k+1|k} > 0$, $\tau_i \geq 0, \ i=7,8,9,10,11,12$ as solutions to the following SDP:
\begin{equation} \label{The complete problem statement-2}
\begin{split} 
& \min_{\bm{P}_{k+1|k}, \tau_7,\tau_8,\tau_9, \tau_{10}, \tau_{11}, \tau_{12}} \hspace{0.2cm} \text{trace}(\bm{P}_{k+1|k}) \\
& \text{subject to} \\
& \bm{P}_{k+1|k} > 0 \\
& \tau_i \geq 0, i = 7,8,9,10,11,12 \\
& \begin{bmatrix}
-\bm{P}_{k+1|k} & & \bm{\Pi}_{k|k} \\ \\
\bm{\Pi}^T_{k|k} & & -\bm{\Psi} (\tau_7,\tau_8,\tau_9, \tau_{10}, \tau_{11}, \tau_{12})
\end{bmatrix} \leq 0 
\end{split}
\end{equation}
where $\bm{\Pi}_{k|k}$ and $\bm{\Psi} (\tau_7,\tau_8,\tau_9, \tau_{10}, \tau_{11}, \tau_{12})$ are given by
\begin{displaymath}  
\begin{split}
\bm{\Pi}&_{k|k}\\
      =& \Big [ \bm{0}_{n} \quad \bm{A}(\hat{\bm{x}}_{k|k}) \bm{E}_{k|k} \quad \bm{I}_n \quad \bm{K}_2 \quad \bm{I}_n \quad \bm{K}_2 \quad \bm{I}_n \Big] \\ 
\bm{\Psi} & (\tau_7, \tau_8,\tau_9, \tau_{10}, \tau_{11}, \tau_{12}) \\
                   =& \hspace{0.1cm} \textnormal{diag} \hspace{0.1cm} (1- \tau_7 - \tau_8 - \tau_9 \gamma_{k|k}^2 \hat{\bm{x}}_{k|k}^\textnormal{T} \hat{\bm{x}}_{k|k} - \tau_{10} r_{A_{k}}^2 \hat{\bm{x}}_{k|k}^\textnormal{T} \hat{\bm{x}}_{k|k}, \\
                    &  \tau_7 \bm{I}_n - \tau_{11} \gamma_{k|k}^2 \bm{E}_{k|k}^\textnormal{T} \bm{E}_{k|k} - \tau_{12} r_{A_{k}}^{2} \bm{E}_{k|k}^\textnormal{T} \bm{E}_{k|k}, \tau_8 \bm{Q}_k^{-1}, \\
                    &  \tau_9 \bm{I}_{n^2}, \tau_{10} \bm{I}_{n}, \tau_{11} \bm{I}_{n^2}, \tau_{12} \bm{I}_{n}).
\end{split}
\end{displaymath}
Furthermore, center of the prediction ellipsoid is given by the predicted state estimate in \eqref{objective-predicted state estimate}. 
\end{theorem}
\begin{IEEEproof}
Utilizing \eqref{SDC form of the nonlinear dynamics} and \eqref{objective-predicted state estimate}, we have
\begin{equation} \label{error at the prediction step}
\begin{split}
& \bm{x}_{k+1} - \hat{\bm{x}}_{k+1|k} \\
                                    & = \ \bm{A}(\bm{x}_k) \bm{x}_k + \bm{w}_k - \bm{A}(\hat{\bm{x}}_{k|k}) \hat{\bm{x}}_{k|k} \\
                                    & = (\bm{A}(\hat{\bm{x}}_{k|k}) + \bm{K}_2 \bm{\Delta}_2 + \bm{R}_{\bm{A}_{2}}) (\hat{\bm{x}}_{k|k} + \bm{E}_{k|k} \bm{z}_{k|k})  \\
                                    &  \quad + \bm{w}_k - \bm{A}(\hat{\bm{x}}_{k|k}) \hat{\bm{x}}_{k|k} \\
                                    & = \bm{A}(\hat{\bm{x}}_{k|k}) \bm{E}_{k|k} \bm{z}_{k|k} + \bm{K}_2 \bm{\Delta}_2 \hat{\bm{x}}_{k|k} + \bm{R}_{\bm{A}_{2}} \hat{\bm{x}}_{k|k}  \\
                                    & \quad  + \bm{K}_2 \bm{\Delta}_2 \bm{E}_{k|k} \bm{z}_{k|k} + \bm{R}_{\bm{A}_{2}} \bm{E}_{k|k} \bm{z}_{k|k} + \bm{w}_k
\end{split} 
\end{equation}
Denote the unknowns in \eqref{error at the prediction step} as
\begin{equation}
\begin{split}
\bm{\Delta}_{7} &= \bm{\Delta}_2 \bm{E}_{k|k} \bm{z}_{k|k}  \\
\bm{\Delta}_{8} &= \bm{R}_{\bm{A}_{2}} \bm{E}_{k|k} \bm{z}_{k|k}  \\
\bm{\Delta}_{9} &= \bm{\Delta}_2 \hat{\bm{x}}_{k|k} \\
\bm{\Delta}_{10} &= \bm{R}_{\bm{A}_{2}} \hat{\bm{x}}_{k|k}. \\
\end{split}
\end{equation}
The rest of the proof can be completed by carrying out steps similar to the ones carried out for the proof of Theorem \ref{Theorem:correction step}.
\end{IEEEproof}
These SDPs in \eqref{The complete problem statement-1} and \eqref{The complete problem statement-2} can be solved efficiently using interior point methods \cite{Vandenberghe_Boyd_1996}. In terms of practical efficiency, interior point methods roughly require 5-50 iterations to solve each SDP with each iteration requiring solution to a least-squares problem of the same size as the original problem \cite{Vandenberghe_Boyd_1996}. The recursive SDC-SMF algorithm for system \eqref{SDC form of the nonlinear dynamics} is summarized in Algorithm \ref{SMF algorithm}.
\begin{algorithm} 
\caption{SDC-SMF Algorithm} 
\label{SMF algorithm}
\begin{algorithmic}[1]
\State (Initialization) Choose a time-horizon $T_f$. Given the initial values $(\hat{\bm{x}}_0, \bm{P}_0)$, set $k = 0$, $\hat{\bm{x}}_{k|k-1} = \hat{\bm{x}}_0$, $\bm{E}_{k|k-1} = \bm{E}_0$ where $ \bm{P}_0 = \bm{E}_0 \bm{E}_0^{\text{T}}$, and $\gamma_{k|k-1} = ||\bm{E}_0||$.
\State Calculate $r_{H_{k}}$ by solving \eqref{r_H}. Find $\bm{P}_{k|k}$ and $\bm{L}_k$ by solving the SDP in \eqref{The complete problem statement-1}. 
\State Calculate $\hat{\bm{x}}_{k|k}$ using \eqref{objective-corrected state estimate}. Also, calculate $\bm{E}_{k|k}$ using $\bm{P}_{k|k} = \bm{E}_{k|k} \bm{E}_{k|k}^\text{T}$ and set $\gamma_{k|k} = || \bm{E}_{k|k} ||$.
\State Calculate $r_{A_{k}}$ by solving \eqref{r_A}. With that, given $\hat{\bm{x}}_{k|k}$, $\bm{E}_{k|k}$, $\gamma_{k|k}$, solve the SDP in \eqref{The complete problem statement-2} to obtain $\bm{P}_{k+1|k}$.
\State Calculate $\hat{\bm{x}}_{k+1|k}$ using \eqref{objective-predicted state estimate}. Set $\bm{E}_{k+1|k}$ using $\bm{P}_{k+1|k} = \bm{E}_{k+1|k} \bm{E}_{k+1|k}^\text{T}$ and $\gamma_{k+1|k} = ||\bm{E}_{k+1|k}||$. 
\State If $k = T_f$ stop. Otherwise, set $k=k+1$ and go to Step 2. 
\end{algorithmic}
\end{algorithm}
\begin{remark} \label{Remark on the conservativeness of the SDC-SMF}
Note that the upper bounds calculated using \eqref{r_A} and \eqref{r_H} are conservative since the points are sampled from the boundary of the ellipsoids, whereas the true state of the system might belong to the interior of these sets. Assumption \ref{Assumption 3 - process and measurement noise ellipsoids} means $||\bm{w}_k|| \leq \sqrt{q}$ and $||\bm{v}_k|| \leq \sqrt{r}$ for all $k \in \mathbb{Z}_\star$. Therefore, higher values of $q$ and $r$ would indicate that the available bounds on the noises are large which would also introduce some degree of conservativeness to the SDC-SMF. 
\end{remark}
Until this point, we have discussed the SDC-SMF for system \eqref{SDC form of the nonlinear dynamics}. Now, let us discuss the application of SDC-SMF to systems with known control inputs, i.e., systems of the form 
\begin{equation}  \label{Systems with a control input}
\begin{split}
\bm{x}_{k+1} & = \bm{f}_d(\bm{x}_k) + \sum_{i=1}^{m} \bm{g}_d (\bm{x}_k) u_{k_{i}} + \bm{w}_k \\
             &  = \bm{A}(\bm{x}_k) \bm{x}_k + \bm{B}(\bm{x}_k) \bm{u}_k + \bm{w}_k \\
\bm{y}_k     & = \bm{h}_d (\bm{x}_k) + \bm{v}_k = \bm{H}(\bm{x}_k) \bm{x}_k + \bm{v}_k
\end{split}
\end{equation} 
where $\bm{u}_k \in \mathbb{R}^m$ is a vector of known control inputs and $\bm{f}_d (\cdot)$, $\bm{h}_d (\cdot)$ again satisfy Assumption \ref{Assumptions on the nonlinear functions}. To be consistent with our earlier formulation, we choose the SDC parameterizations given in \eqref{SDC form calculation-2} and state the following assumption regarding the state dynamics of system \eqref{Systems with a control input}.
\begin{assumption} \label{Assumption on state trajectory with control input}
There exist compact sets $\mathbb{D}_{u_{0}}, \mathbb{D}_u \subset \mathbb{R}^n$, $\mathbb{U} \subset \mathbb{R}^m$, and $\epsilon_u > 0$ such that $\bm{x}_0 \in \mathbb{D}_{u_{0}}$ and $\bm{u}_k \in \mathbb{U}$ together imply
\begin{displaymath}
\bm{x}_k + \epsilon_u \mathcal{B}(\bm{x}_k) \subset \mathbb{D}_u, \quad \forall k \in \mathbb{Z}_{\star} .
\end{displaymath}
\end{assumption} 
The implication of the above assumption is similar to that of Assumption \ref{State dynamics compactness assumption}, i.e., the system \eqref{Systems with a control input} evolves within a compact set $\mathbb{D}_u$ which is not necessarily small. 
Then, the filtering problem at the correction step is as in Theorem \ref{Theorem:correction step} with system \eqref{SDC form of the nonlinear dynamics} replaced by system \eqref{Systems with a control input} and $\mathbb{D}_0$ in Assumption \ref{Assumption 3 - initial conditions} replaced by $\mathbb{D}_{u_{0}}$. However, the SDP for the prediction step would have to be modified due to the control inputs acting through the state dependent control matrix. To this end, similar to the matrix Taylor expansion of $\bm{A}(\bm{x}_k)$ in \eqref{Expansion of A(x)}, let us expand $\bm{B}(\bm{x}_k)$ as
\begin{equation}
\begin{split}
\bm{B}({\bm{x}}_k) =& \ \bm{B}(\hat{\bm{x}}_{k|k}) + \bm{K}_3 (\hat{\bm{x}}_{k|k}) \bm{\Delta}_3 (\bm{\xi}_{k|k}) \\
                    &+ \bm{R}_{\bm{B}_{2}} (\hat{\bm{x}}_{k|k}, \bm{x}_{k})
\end{split}
\end{equation}
where $\bm{K}_3 (\hat{\bm{x}}_{k|k}) = \mathbb{D}_{\bm{x}^\text{T}} \bm{B}(\hat{\bm{x}}_{k|k})$, $\bm{\Delta}_3 (\bm{\xi}_{k|k}) = \left(\bm{\xi}_{k|k} \otimes \bm{I}_m \right)$ with $\bm{\xi}_{k|k}$ as in \eqref{Expansion of A(x)}. Again, similar to \eqref{r_A}- \eqref{r_H}, let us calculate the upper bound on the norm of remainder $\bm{R}_{\bm{B}_{2}} (\hat{\bm{x}}_{k|k}, \bm{x}_{k})$ as
\begin{equation}  \label{r_B}
\begin{split}
r_{B_{k}} =  \max_{i = 1,2,...,N} || \bm{R}_{\bm{B}_{2}} (\hat{\bm{x}}_{k|k}, \hat{\bm{x}}_{k|k} + \bm{E}_{k|k} \bm{z}_{{k|k}_{i}})||
\end{split}
\end{equation}
where $||\bm{z}_{{k|k}_{i}}|| = 1, \ i = 1,2,...,N$. Finally, the next result summarizes the filtering problem at the prediction step for systems with state dynamics as in \eqref{Systems with a control input} where we have dropped the argument of $\bm{K}_3 (\cdot)$.
\begin{corollary}
Consider system \eqref{Systems with a control input} under Assumption \ref{Assumption 3 - process and measurement noise ellipsoids} with the current state 
$\bm{x}_k \in \mathcal{E} (\hat{\bm{x}}_{k|k}, \bm{P}_{k|k})$ and $\bm{w}_k \in \mathcal{E} (\bm{0}_n, \bm{Q}_k)$. Then, the successor state $\bm{x}_{k+1}$ belongs to the optimal prediction ellipsoid $\mathcal{E} (\hat{\bm{x}}_{k+1|k}, \bm{P}_{k+1|k})$, if there exist $\bm{P}_{k+1|k} > 0$, $\tau_1,\tau_2,\tau_3, \tau_4, \tau_5, \tau_6, \tau_7,\tau_8 \geq 0$ as solutions to the following SDP: 
\begin{equation*} 
\begin{split} 
& \min_{\bm{P}_{k+1|k}, \tau_1,\tau_2,\tau_3, \tau_4, \tau_5, \tau_6, \tau_7,\tau_8} \hspace{0.2cm} \textnormal{trace}(\bm{P}_{k+1|k}) \\
& \text{subject to} \\
& \bm{P}_{k+1|k} > 0 \\
& \tau_i \geq 0, \ i=1,2,3,4,5,6,7,8 \\
& \begin{bmatrix}
-\bm{P}_{k+1|k} & & \bm{\Pi}_{k|k} \\ \\
\bm{\Pi}^T_{k|k} & & -\bm{\Psi} (\tau_1,\tau_2,\tau_3, \tau_4, \tau_5, \tau_6, \tau_7,\tau_8)
\end{bmatrix} \leq 0
\end{split}
\end{equation*}
where $\bm{\Pi}_{k|k}$ and $\bm{\Psi} (\tau_1,\tau_2,\tau_3, \tau_4, \tau_5, \tau_6, \tau_7,\tau_8)$ are given by
\begin{equation*} 
\begin{split}
\bm{\Pi}&_{k|k} \\
       =& \Big [
\bm{0}_{n} \quad \bm{A}(\hat{\bm{x}}_{k|k}) \bm{E}_{k|k} \quad \bm{I}_n \quad \bm{K}_2 \quad \bm{I}_n \quad \bm{K}_2 \quad \bm{I}_n \quad
\bm{K}_3 \quad \bm{I}_n \Big] \\ 
\bm{\Psi} & (\tau_1,\tau_2,\tau_3, \tau_4, \tau_5, \tau_6, \tau_7,\tau_8) \\
            =& \hspace{0.1cm} \textnormal{diag} \hspace{0.1cm} (1- \tau_1 - \tau_2 - \tau_3 \gamma_{k|k}^2 \hat{\bm{x}}_{k|k}^\textnormal{T} \hat{\bm{x}}_{k|k} - \tau_4 r_A^2 \hat{\bm{x}}_{k|k}^\textnormal{T} \hat{\bm{x}}_{k|k} \\
                                   & - \tau_7 \gamma_{k|k}^2  \bm{u}_k^\textnormal{T} \bm{u}_k - \tau_8 r_{B_{k}}^2 \bm{u}_k^\textnormal{T} \bm{u}_k, \tau_1 \bm{I}_n - \tau_5 \gamma_{k|k}^2 \bm{E}_{k|k}^\textnormal{T} \bm{E}_{k|k} \\
& - \tau_6 r_{A_{k}}^{2} \bm{E}_{k|k}^\textnormal{T} \bm{E}_{k|k}, \tau_2 \bm{Q}_k^{-1},  \tau_3 \bm{I}_{n^2}, \tau_4 \bm{I}_{n}, \tau_5 \bm{I}_{n^2}, \\
& \tau_6 \bm{I}_{n}, \tau_7 \bm{I}_{mn}, \tau_8 \bm{I}_{n}).
\end{split}
\end{equation*}
Furthermore, the center of the prediction ellipsoid is given by the predicted state estimate 
\begin{equation}
\hat{\bm{x}}_{k+1|k} = \bm{A}(\hat{\bm{x}}_{k|k}) \hat{\bm{x}}_{k|k} + \bm{B}(\hat{\bm{x}}_{k|k}) \bm{u}_k.
\end{equation}
\end{corollary}
\begin{IEEEproof}
Follows from that of Theorem \ref{Theorem:prediction step} and is omitted.
\end{IEEEproof}
\section{Theoretical properties of SDC-SMF} \label{Theoretical properties of SDC-SMF}
In this section, we provide an elaborate sketch of the proof of the theoretical properties satisfied by the proposed SDC-SMF (which are similar to the observer properties described in Definition 3.1 in \cite{Shamma_Tu_1997}). To this end, we utilize the approach outlined in Sections IV, V in \cite{Shamma_Tu_1997} and adopt the symbols used to represent some variables in \cite{Shamma_Tu_1997} so that it is easy to draw parallels between the results given here and the results in \cite{Shamma_Tu_1997}.
Further, for a sequence $\bm{x} = \{ \bm{x}_k  \}_{k \in \mathbb{Z}_\star}$ with $\bm{x}_k \in \mathbb{R}^n$ and $k \in \mathbb{Z}_\star$, we denote $|| \bm{x} ||_{l^\infty} = \sup_{k \in \mathbb{Z}_\star} ||\bm{x}_k||$. Due to the differences in the notations, simple modifications have to be introduced for Definition 3.1 in \cite{Shamma_Tu_1997} and it is understood that those changes have already been carried out. 

\begin{remark}
Initialization step of the Algorithm 4.1 in \cite{Shamma_Tu_1997} is similar to the initial correction step for the SDC-SMF at $k=0$. Then, the the Algorithm 4.1 in \cite{Shamma_Tu_1997} employs a one-step estimation wherein the correction and prediction are combined into one single step. For the SDC-SMF, we have two distinct steps for correction and prediction. However, for the analysis shown here, we would only consider the correction step with the corrected state estimate explicitly. The prediction step is only considered implicitly in the subsequent analysis. With this approach, we show that the corrected state estimate satisfies properties similar to the ones given in Definition 3.1 in \cite{Shamma_Tu_1997}. Then, the same applies for the predicted state estimate under the conditions/assumptions described in the sequel.
\end{remark}

Consider the simplified version of system \eqref{SDC form of the nonlinear dynamics} given by
\begin{equation} \label{SDC form of the simplified nonlinear system}
\begin{split}
\bm{x}_{k+1} &= \bm{A}(\bm{x}_k) \bm{x}_k + \bm{w}_k \\
\bm{y}_k &= \bm{H} \bm{x}_k + \bm{v}_k
\end{split}
\end{equation}
where the state-dependent matrix $\bm{H}(\bm{x}_k)$ is replaced by the constant matrix $\bm{H}$. This obviously introduces some loss of generality (which is remarked by the authors in \cite{Shamma_Tu_1997} as well), but is crucial for establishing the theoretical properties, as shown in the sequel. Now, let (i) Assumption \ref{State dynamics compactness assumption} hold for the state dynamics of system \eqref{SDC form of the simplified nonlinear system}; (ii) Assumption \ref{Assumption: initial conditions, noises, and trajectory for the true system} hold for system \eqref{SDC form of the simplified nonlinear system}; (iii) Assumption \ref{Observability related assumption} hold with system \eqref{SDC form of the nonlinear dynamics} replaced by system \eqref{SDC form of the simplified nonlinear system} and $\bm{H}(\bm{x}_{(\cdot)})$ replaced by $\bm{H}$. With that, let us implement the proposed SDC-SMF for system \eqref{SDC form of the simplified nonlinear system}. Note that Assumptions 3.1 and 5.2.1 in \cite{Shamma_Tu_1997} are replaced by our Assumption \ref{Assumption 3 - process and measurement noise ellipsoids}. Under our Assumption \ref{Assumption 3 - process and measurement noise ellipsoids}, we have $||\bm{w}||_{l^\infty} \leq \sqrt{q}$ and $||\bm{v}||_{l^\infty} \leq \sqrt{r}$.  

Before discussing the theoretical properties of the SDC-SMF for system \eqref{SDC form of the simplified nonlinear system}, we give the next two assertions (Claims \ref{Claim on the observability along the estimate trajectory} and \ref{Claim on the uniform boundedness of the term d_f_k}), under our above assumptions. First, we adopt the following claim from \cite[Section V]{Shamma_Tu_1997} which is asserted to hold due to the time-invariance and compactness assumptions. 

\begin{claim}  \label{Claim on the observability along the estimate trajectory}
Let $\alpha>0$ and $\epsilon_2>0$ be such that $\forall k \in \mathbb{Z}_\star$
\begin{itemize}
\item $||\bm{A}(\bm{x})|| \leq \alpha, \ \forall \bm{x} \in \mathbb{D} $
\item $||\bm{x} - \hat{\bm{x}}||_{l^\infty} \leq \epsilon_2$ with $\hat{\bm{x}} = \{ \hat{\bm{x}}_{k|k}  \}_{k \in \mathbb{Z}_\star}$ implies
\begin{equation*}
\mu_1^o \bm{I}_n \leq \hat{\mathcal{O}}_{k,k+N_o-1}^\text{T} \hat{\mathcal{O}}_{k,k+N_o-1} \leq \mu_2^o \bm{I}_n
\end{equation*} 
where $\mu_1^o (\mu_1, \mu_2) > 0$, $\mu_2^o (\mu_1, \mu_2) > 0$, and
\begin{equation*}
\hat{\mathcal{O}}_{k,k+s} = 
\begin{bmatrix}
\bm{H} \\
\bm{H} \hat{\bm{\phi}}_{k+1,k} \\
\vdots \\
\bm{H} \hat{\bm{\phi}}_{k+s,k}
\end{bmatrix}
\end{equation*}
with 
\begin{equation*}
\begin{split}
\hat{\bm{\phi}}_{k+s,k} = & \bm{A}(\hat{\bm{x}}_{k+s-1|k+s-1}) \ \bm{A}(\hat{\bm{x}}_{k+s-2|k+s-2}) \\
                          & \cdots \bm{A}(\hat{\bm{x}}_{k|k}) \\
\end{split}
\end{equation*} 
defined along the corrected state estimate trajectory for any $ s \in \mathbb{Z}_\star \backslash \{0\}$.
\end{itemize}
\end{claim}

Using the matrix Taylor expansion of $\bm{A}(\bm{x}_k)$, we have the state dynamics of the form (cf., \eqref{dynamics utilized for filter design})
\begin{equation*}
\begin{split}
\bm{x}_{k+1} = \ \bm{A}(\hat{\bm{x}}_{k|k}) \bm{x}_k + \bm{K}_2 \bm{\Delta}_2 \bm{x}_k 
              + \bm{R}_{\bm{A}_{2}} \bm{x}_k + \bm{w}_k
\end{split}
\end{equation*}
where $\bm{\Delta}_2 = (\bm{x}_k - \hat{\bm{x}}_{k|k}) \otimes \bm{I}_n$ and we define 
\begin{equation*}
\begin{split}
\bm{d}_{f_{k}} &= \bm{K}_2 \bm{\Delta}_2 \bm{x}_k + \bm{R}_{\bm{A}_{2}}  \bm{x}_k \\
\mathbb{E}_k &= \{ \bm{\nu} : \hat{\bm{x}}_{k|k} +  \bm{\nu} \in \mathcal{E}(\hat{\bm{x}}_{k|k}, \bm{P}_{k|k})  \} \\
\rho_k &= \sup_{\bm{\nu} \in \mathbb{E}_k} ||\bm{\nu}||.
\end{split}
\end{equation*}

Next, the following claim is related to the norm of the term $\bm{d}_{f_{k}}$ where $\epsilon_1$ and $\mathbb{D}$ are as in Assumption \ref{State dynamics compactness assumption}.

\begin{claim} \label{Claim on the uniform boundedness of the term d_f_k}
Define $\bar{\epsilon} = \max \{\epsilon_1, \epsilon_2 \}$. Also, define a compact subset $\bar{\mathbb{D}} \subset \mathbb{R}^n$ such that $\bar{\mathbb{D}} = \mathbb{D}$ if $\bar{\epsilon} = \epsilon_1$, otherwise $\bar{\mathbb{D}} \supseteq \mathbb{D}$ with $d_H(\mathbb{D},\bar{\mathbb{D}}) \leq \bar{\epsilon}$ where $d_H(\cdot, \cdot)$ is the Hausdorff distance. Let there exist $\bar{a} > 0$ such that $||\bm{A}(\bm{x}_1) - \bm{A}(\bm{x}_2)|| \leq \bar{a} ||\bm{x}_1 - \bm{x}_2||$ for all $\bm{x}_1, \bm{x}_2 \in \bar{\mathbb{D}}$. Then, 
\begin{equation*}
||\bm{d}_{f_{k}}|| \leq \delta \rho_k
\end{equation*}
for some $\delta>0$.
\end{claim}
\begin{IEEEproof}
The remainder of the matrix Taylor expansion can be expressed as in \eqref{R_A expression for the remainder norm max calculation} with $\bm{K}_2 \equiv \bm{K}_2(\hat{\bm{x}}_{k|k})$.
With the assertion in Claim \ref{Claim on the observability along the estimate trajectory} and the definition of the set $\bar{\mathbb{D}}$, we have $\bm{x}_k, \hat{\bm{x}}_{k|k} \in \bar{\mathbb{D}}$. Thus, under the assumption that $||\bm{A}(\bm{x}_1) - \bm{A}(\bm{x}_2)|| \leq \bar{a} ||\bm{x}_1 - \bm{x}_2||$, we have
\begin{equation*}
|| \bm{R}_{\bm{A}_{2}}|| \leq \bar{a} ||\bm{x}_k - \hat{\bm{x}}_{k|k}|| + ||\bm{K}_2|| \ ||\bm{x}_k - \hat{\bm{x}}_{k|k}||
\end{equation*}
since $\bm{\Delta}_2 = (\bm{x}_k - \hat{\bm{x}}_{k|k}) \otimes \bm{I}_n$. Also, $||\bm{K}_2|| \leq k_2$ for some $k_2 > 0$ holds due to the continuity of $\bm{K}_2$ and compactness of $\bar{\mathbb{D}}$. Collecting all these, we deduce
\begin{equation*}
|| \bm{R}_{\bm{A}_{2}}|| \leq (\bar{a} + k_2) ||\bm{x}_k - \hat{\bm{x}}_{k|k}|| \leq \alpha_r \rho_k
\end{equation*}
with some $\alpha_r > 0$. Also, $||\bm{x}_k|| \leq \alpha_x$ with some $\alpha_x > 0$ holds due to the compactness of $\mathbb{D}$. Therefore,
\begin{equation*}
\begin{split}
||\bm{d}_{f_{k}}|| & \leq ||\bm{K}_2|| \ ||\bm{x}_k - \hat{\bm{x}}_{k|k}|| \ ||\bm{x}_k||  + || \bm{R}_{\bm{A}_{2}}|| \ ||\bm{x}_k|| \\
                   & \leq k_2 \alpha_x \rho_k + \alpha_r \alpha_x \rho_k.
\end{split}
\end{equation*}
Combining all the above results, we conclude that there is a constant $\delta > 0$ such that
\begin{equation*}
||\bm{d}_{f_{k}} || \leq \delta \rho_k
\end{equation*} 
holds $\forall k \in \mathbb{Z}_\star$. 
\end{IEEEproof}
\begin{remark}
Note that we have shown that the norm of the remainder term remains uniformly bounded under the Lipschitz continuity assumption on the matrix valued function $\bm{A}(\cdot)$. We stress that this assumption would hold due the continuous differentiability of the function and compactness of the sets. Furthermore, note that the above bound on the remainder term is developed using the methodology in Section \ref{Preliminaries and Problem Formulation} to calculate $r_{A_{k}}$ at each time step. Thus, $r_{A_{k}}$ would implicitly obey the above bound as well. 
\end{remark}

We are now ready to establish the theoretical properties of the SDC-SMF for system \eqref{SDC form of the simplified nonlinear system}. To this end, we first show that the SDC-SMF is nondivergent in the presence of the process and measurement noises and is unbiased and asymptotically convergent in the absence of the noises. 

\subsection{Nondivergence for \texorpdfstring{$\bm{w}_k \neq \bm{0}_n$}{TEXT} and \texorpdfstring{$\bm{v}_k \neq \bm{0}_p$}{TEXT}} 
First, let us redefine the `false' system in \cite[Section IV.B]{Shamma_Tu_1997}. Consider the following system 
\begin{equation} \label{false system-1}
\begin{split}
\bm{x}_{f_{{k+1}}} &= \bm{A}(\hat{\bm{x}}_{k|k}) \bm{x}_{f_{k}} + \bm{d}_{f_{k}} + \bm{u}_{f_{k}} + \bm{w}_k \\
\bm{x}_{f_{0}}     &= \bm{0}_n \\
\bm{y}_{f_{k}}     &= \bm{H} \bm{x}_{f_{k}} + \bm{v}_k
\end{split}
\end{equation}
where $\bm{x}_{f_{k}} = \bm{x}_k - \hat{\bm{x}}_{k|k}$, $\bm{u}_{f_{k}} = \bm{A}(\hat{\bm{x}}_{k|k})\hat{\bm{x}}_{k|k} - \hat{\bm{x}}_{k+1|k+1}$. Note that this system is non-causal as in \cite{Shamma_Tu_1997} and we have implicitly utilized the predicted state estimate in $\bm{u}_{f_{k}}$ as $\hat{\bm{x}}_{k+1|k} = \bm{A}(\hat{\bm{x}}_{k|k})\hat{\bm{x}}_{k|k}$. Next, we state an important result that is subsequently utilized to show that the SDC-SMF is nondivergent for the case under consideration.
\begin{proposition} \label{Proposition regarding the rho_k bound-1}
Given any $j \in \mathbb{Z}_\star$, let 
\begin{equation*}
\begin{split}
g_1 &= \max \{ 1, \theta^j (\alpha + \delta)^j \} \\
g_2 &= \sum_{k=1}^{j} \theta^k (\alpha + \delta)^{k-1} \\
g_3 &= \sum_{k=1}^{j} \bar{l} \ \theta^{k-1} (\alpha + \delta)^{k-1}
\end{split}
\end{equation*}
where $\theta = (1 + \bar{l} \ ||\bm{H}||)$ with $||\bm{L}_k|| \leq \bar{l}, \ \forall k \in [1,j]$ for some $\bar{l} > 0$. Then, 
\begin{equation*}
\max_{0 \leq k \leq j} \rho_k \leq g_1 \rho_0 + g_2 ||\bm{w}||_{l^\infty} + + g_3 ||\bm{v}||_{l^\infty}.
\end{equation*}
\end{proposition} 

\begin{IEEEproof}
For any $k \in \mathbb{Z}_\star\backslash\{0\}$, we have 
\begin{equation*}
\bm{x}_k - \hat{\bm{x}}_{k|k} = \bm{x}_k - \hat{\bm{x}}_{k|k-1} - \bm{L}_k \bm{H} \ (\bm{x}_k - \hat{\bm{x}}_{k|k-1}) - \bm{L}_k \bm{v}_k
\end{equation*}
where
\begin{equation*}
\begin{split}
\bm{x}_{k} &= \bm{A}(\hat{\bm{x}}_{k-1|k-1}) \bm{x}_{k-1} + \bm{d}_{f_{k-1}} + \bm{w}_{k-1} \\
\hat{\bm{x}}_{k|k-1} &= \bm{A}(\hat{\bm{x}}_{k-1|k-1}) \hat{\bm{x}}_{k-1|k-1} .
\end{split}
\end{equation*}
Therefore, we can write
\begin{equation*}
\begin{split}
& \bm{x}_k - \hat{\bm{x}}_{k|k} \\
& = \bm{A}(\hat{\bm{x}}_{k-1|k-1}) \bm{x}_{k-1} + \bm{d}_{f_{k-1}} + \bm{w}_{k-1} \\
& \quad - \bm{A}(\hat{\bm{x}}_{k-1|k-1}) \hat{\bm{x}}_{k-1|k-1} - \bm{L}_k \bm{H} \Big( \bm{A}(\hat{\bm{x}}_{k-1|k-1}) \bm{x}_{k-1} \\
& \quad + \bm{d}_{f_{k-1}} + \bm{w}_{k-1} - \bm{A}(\hat{\bm{x}}_{k-1|k-1}) \hat{\bm{x}}_{k-1|k-1} \Big)  - \bm{L}_k \bm{v}_k \\
& = \ \bm{A}(\hat{\bm{x}}_{k-1|k-1}) \left( \bm{x}_{k-1} - \hat{\bm{x}}_{k-1|k-1} \right) + \bm{d}_{f_{k-1}} + \bm{w}_{k-1} \\
& \quad - \bm{L}_k \bm{H} \left( \bm{A}(\hat{\bm{x}}_{k-1|k-1}) \Big( \bm{x}_{k-1} - \hat{\bm{x}}_{k-1|k-1} \right) + \bm{d}_{f_{k-1}} \\
& \quad + \bm{w}_{k-1} \Big) - \bm{L}_k \bm{v}_k .
\end{split}
\end{equation*}
Let $\bar{l} > 0$ be such that $||\bm{L}_k|| \leq \bar{l}, \ \forall k \in [1,j]$. Hence, we derive
\begin{equation*}
\begin{split}
\rho_k =& \max_{\bm{x}_k \in \mathcal{E}(\hat{\bm{x}}_{k|k}, \bm{P}_{k|k})} ||\bm{x}_k - \hat{\bm{x}}_{k|k}|| \\
\leq & \ \alpha \rho_{k-1} + \delta \rho_{k-1} + ||\bm{w}||_{l^\infty} \\
& + ||\bm{L}_k|| \ ||\bm{H}|| (\alpha \rho_{k-1} + \delta \rho_{k-1} + ||\bm{w}||_{l^\infty}) \\
& + ||\bm{L}_k|| \ ||\bm{v}||_{l^\infty} \\
\leq & (1 + \bar{l} \ ||\bm{H}||) \left( (\alpha + \delta) \rho_{k-1} + ||\bm{w}||_{l^\infty} \right) + \bar{l} \ ||\bm{v}||_{l^\infty} .
\end{split}
\end{equation*}
Carrying out these calculations recursively yields
\begin{equation*}
\begin{split}
\rho_j \leq & \theta^j (\alpha + \delta)^j \rho_0 + \sum_{k=1}^{j} \theta^k (\alpha + \delta)^{k-1} ||\bm{w}||_{l^\infty} \\
            & + \sum_{k=1}^{j} \bar{l} \ \theta^{k-1} (\alpha + \delta)^{k-1} ||\bm{v}||_{l^\infty} 
\end{split}
\end{equation*}
where $\theta = (1 + \bar{l} \ ||\bm{H}||)$. Then, collecting all the required bounds leads to the desired result.
\end{IEEEproof}
\begin{remark}
Note that we have used a uniform bound $||\bm{L}_k|| \leq \bar{l}$ for the filter gain. This is guaranteed to hold as the filter gain is a solution to a convex optimization problem (namely, SDP) at each time step. 
\end{remark}

Let $\epsilon^\star = \min \{\epsilon_1, \epsilon_2\}$. We need to show that for $k \in [0,N_o-1]$
\begin{equation*}
||\bm{x}_k - \hat{\bm{x}}_{k|k}|| \leq \epsilon^\star .
\end{equation*} 
Then,
\begin{equation} \label{Gramian bounds for 0 to N-1}
\mu_1^o \bm{I}_n \leq \hat{\mathcal{O}}_{0,N_o-1}^\text{T} \hat{\mathcal{O}}_{0,N_o-1} \leq \mu_2^o \bm{I}_n.
\end{equation}
Making straightforward modifications to the result in Proposition \ref{Proposition regarding the rho_k bound-1}, we can assure
\begin{equation*}
\max_{0 \leq k \leq N_o-1} \rho_k \leq \epsilon^\star
\end{equation*}
whenever $\bm{x}_0 \in \mathbb{D}_0$, $\rho_0 \leq \bar{\rho}_1$, $||\bm{w}||_{l^\infty} \leq \bar{d}$, and $||\bm{v}||_{l^\infty} \leq \bar{n}$ where
\begin{equation*}
\begin{split}
\bar{\rho}_1 &= \min \left\{\frac{\epsilon^\star}{3}, \frac{(\epsilon^\star/3)}{\theta^{N_o-1} (\alpha + \delta)^{N_o-1}}  \right\} \\
\bar{d}    &= \frac{(\epsilon^\star/3)}{\sum_{k=1}^{N_o-1} \theta^{k} (\alpha + \delta)^{k-1}} \\
\bar{n}    &= \frac{(\epsilon^\star/3)}{\sum_{k=1}^{N_o-1} \bar{l} \ \theta^{k-1} (\alpha + \delta)^{k-1}}.
\end{split}
\end{equation*}
This, in turn, implies \eqref{Gramian bounds for 0 to N-1}. 

Next, let us implement the gramian-based observer for the `false' system \eqref{false system-1}, as in \cite{Shamma_Tu_1997}. Doing so, we have
\begin{equation*}
\begin{split}
&||\bm{x} - \hat{\bm{x}}_{N_o|N_o} - \hat{\bm{x}}_{g_{N_o}}|| \\
& \leq \beta_1 (||\bm{w}||_{l^\infty} + \max_{0\leq k \leq N_o-1} ||\bm{d}_{f_{k}}||) + \beta_2 ||\bm{v}||_{l^\infty} \\ 
& \leq \beta_1 (||\bm{w}||_{l^\infty} + \delta \max_{0\leq k \leq N_o-1} \rho_k) + \beta_2 ||\bm{v}||_{l^\infty} \\
\end{split}
\end{equation*}
for any $\bm{x} \in \mathcal{E}(\hat{\bm{x}}_{N_o|N_o}, \bm{P}_{N_o|N_o})$ and with $\beta_1, \beta_2 > 0$. Note that the above bound holds for $\bm{x} = \hat{\bm{x}}_{N_o|N_o}$. With this, we derive
\begin{equation*}
\begin{split}
\rho_{N_{o}} &= \max_{\bm{x} \in \mathcal{E}(\hat{\bm{x}}_{N_o|N_o}, \bm{P}_{N_o|N_o})} ||\bm{x} - \hat{\bm{x}}_{N_o|N_o}||  \\
             & \leq ||\bm{x} - \hat{\bm{x}}_{N_o|N_o} - \hat{\bm{x}}_{g_{N_{o}}}|| + ||\hat{\bm{x}}_{g_{N_{o}}}|| \\
             & \leq 2 \beta_1 (||\bm{w}||_{l^\infty} + \delta \max_{0\leq k \leq N_o-1} \rho_k) + 2 \beta_2 ||\bm{v}||_{l^\infty} .
\end{split}
\end{equation*} 

Utilizing the result in Proposition \ref{Proposition regarding the rho_k bound-1} with $j=N_o-1$, we have
\begin{equation*}
\begin{split}
\rho_{N_{o}} \leq & \ 2 \beta_1 \left( ||\bm{w}||_{l^\infty} + \delta (g_1 \rho_0 + g_2 ||\bm{w}||_{l^\infty} + g_3 ||\bm{v}||_{l^\infty}) \right) \\
&+ 2 \beta_2 ||\bm{v}||_{l^\infty}
\end{split}
\end{equation*}
which upon rearranging becomes
\begin{equation*}
\rho_{N_{o}} \leq c_1 \rho_0 + c_2 ||\bm{w}||_{l^\infty} + c_3 ||\bm{v}||_{l^\infty}
\end{equation*}
where $c_1 = 2 \beta_1 \delta g_1$, $c_2 = 2 \beta_1 (1 + \delta g_2)$, and $c_3 = 2(\beta_2 + \beta_1 \delta g_3)$. Thus,
\begin{equation*}
\begin{split}
\rho_0 & \leq \frac{\bar{\rho}_1}{(c_1 + c_2 + c_3) (c_1 + c_2 + c_3 + 1)} \\
||\bm{w}||_{l^\infty} & \leq \min \left\{ \bar{d}, \frac{\bar{\rho}_1}{(c_1 + c_2 + c_3) (c_1 + c_2 + c_3 + 1)} \right\} \\
||\bm{v}||_{l^\infty} & \leq \min \left\{ \bar{n}, \frac{\bar{\rho}_1}{(c_1 + c_2 + c_3) (c_1 + c_2 + c_3 + 1)} \right\} 
\end{split}
\end{equation*}
together imply 
\begin{equation*}
\rho_{N_{o}} \leq \frac{\bar{\rho}_1}{(c_1 + c_2 + c_3 + 1)}
\end{equation*}
which is similar to the result (12) in \cite{Shamma_Tu_1997}. Therefore, the rest of the proof of uniform boundedness of $\rho_k$ and nondivergence of the corrected state estimate follows from arguments similar to the ones outlined in \cite{Shamma_Tu_1997}.

\subsection{Unbiased and asymptotically convergent for \texorpdfstring{$\bm{w}_k = \bm{0}_n$}{TEXT} and \texorpdfstring{$\bm{v}_k = \bm{0}_p$}{TEXT}}
In this case, the SDC-SMF is clearly unbiased for $\mathcal{E}(\hat{\bm{x}}_0, \bm{P}_0) = \bm{x}_0$. Next, let us redefine the `false' system of Proposition 4.1 in \cite{Shamma_Tu_1997}. Consider the following system 
\begin{equation} \label{false system-2}
\begin{split}
\bm{x}_{f_{{k+1}}} &= \bm{A}(\hat{\bm{x}}_{k|k}) \bm{x}_{f_{k}} + \bm{d}_{f_{k}} + \bm{u}_{f_{k}} \\
\bm{y}_{f_{k}}     &= \bm{H} \bm{x}_{f_{k}} 
\end{split}
\end{equation}
where $\bm{x}_{f_{k}} = \bm{x}_k - \hat{\bm{x}}_{k|k}$, $\bm{u}_{f_{k}} = \bm{A}(\hat{\bm{x}}_{k|k})\hat{\bm{x}}_{k|k} - \hat{\bm{x}}_{k+1|k+1}$. This system is obviously similar to the earlier `false' system \eqref{false system-1}. Now, we state a result similar to the one in Proposition \ref{Proposition regarding the rho_k bound-1}.

\begin{proposition} \label{Proposition regarding the rho_k bound-2}
Given any $j \in \mathbb{Z}_\star$, let 
\begin{equation*}
\begin{split}
g &= \max \left\{ 1, \theta^j (\alpha + \delta)^j  \right\} \\
\end{split}
\end{equation*}
where $\theta = (1 + \bar{l} \ ||\bm{H}||)$ with $||\bm{L}_k|| \leq \bar{l}, \forall k \in [1,j]$. Then,
\begin{equation*}
\max_{0 \leq k \leq j} \rho_k \leq g \rho_0 .
\end{equation*}
\end{proposition}

\begin{IEEEproof}
For any $k \in \mathbb{Z}_\star\backslash\{0\}$, we have 
\begin{equation*}
\begin{split}
&\bm{x}_k - \hat{\bm{x}}_{k|k}  \\
& = \ \bm{A}(\hat{\bm{x}}_{k-1|k-1}) \left( \bm{x}_{k-1} - \hat{\bm{x}}_{k-1|k-1} \right) + \bm{d}_{f_{k-1}} \\
& \quad - \bm{L}_k \bm{H} \left( \bm{A}(\hat{\bm{x}}_{k-1|k-1}) \left( \bm{x}_{k-1} - \hat{\bm{x}}_{k-1|k-1} \right) + \bm{d}_{f_{k-1}} \right).
\end{split}
\end{equation*}
As earlier, let $\bar{l} > 0$ be such that $||\bm{L}_k|| \leq \bar{l}, \ \forall k \in [1,j]$. With that, the above expression implies
\begin{equation*}
\begin{split}
\rho_k =& \max_{\bm{x}_k \in \mathcal{E}(\hat{\bm{x}}_{k|k}, \bm{P}_{k|k})} ||\bm{x}_k - \hat{\bm{x}}_{k|k}|| \\
       &\leq \alpha \rho_{k-1} + \delta \rho_{k-1} + ||\bm{L}_k|| \ ||\bm{H}|| (\alpha \rho_{k-1} + \delta \rho_{k-1} ) \\
       &\leq (1 + \bar{l} \ ||\bm{H}||) (\alpha + \delta) \rho_{k-1} .
\end{split}
\end{equation*}
Proceeding recursively for $k=1,2,\dots,j$ leads to the desired result.
\end{IEEEproof}

Same as earlier, we need to show that for $k \in [0,N_o-1]$
\begin{equation*}
||\bm{x}_k - \hat{\bm{x}}_{k|k}|| \leq \epsilon^\star .
\end{equation*} 
To this end, using the result in Proposition \ref{Proposition regarding the rho_k bound-2}, we have
\begin{equation*}
\max_{0 \leq k \leq N_o-1} \rho_k \leq \epsilon^\star
\end{equation*}
whenever $\bm{x}_0 \in \mathbb{D}_0$, $\rho_0 \leq \bar{\rho}_2$ where
\begin{equation*}
\bar{\rho}_2 = \min \left\{ \epsilon^\star, \frac{\epsilon^\star}{\theta^{N_o-1} (\alpha + \delta)^{N_o-1}}  \right\} 
\end{equation*}
which, in turn, implies \eqref{Gramian bounds for 0 to N-1}. 

Next, we implement the gramian-based observer, as in \cite{Shamma_Tu_1997}, for the `false' system \eqref{false system-2} and derive 
\begin{equation*}
\begin{split}
& ||\bm{x} - \hat{\bm{x}}_{N_o|N_o} - \hat{\bm{x}}_{g_{N_o}}|| \\
& \leq \beta \max_{0\leq k \leq N_o-1} ||\bm{d}_{f_{k}}|| \leq \beta \delta \max_{0\leq k \leq N_o-1} \rho_k 
\end{split}
\end{equation*}
for any $\bm{x} \in \mathcal{E}(\hat{\bm{x}}_{N_o|N_o}, \bm{P}_{N_o|N_o})$ and with $\beta > 0$. Therefore, utilizing the result in Proposition \ref{Proposition regarding the rho_k bound-2} with $j=N_o-1$, we have
\begin{equation*}
\begin{split}
\rho_{N_{o}} &= \max_{\bm{x} \in \mathcal{E}(\hat{\bm{x}}_{N_o|N_o}, \bm{P}_{N_o|N_o})} ||\bm{x} - \hat{\bm{x}}_{N_o|N_o}|| \\
             & \leq 2 \beta \delta \max_{0\leq k \leq N_o-1} \rho_k   \leq 2 \beta \delta g \rho_0 .
\end{split}
\end{equation*} 
Then, for 
\begin{equation*}
\beta \leq \frac{\lambda}{2 \delta g}, \ \lambda \in (0,1), \ \rho_0 \leq \bar{\rho}_2 ,
\end{equation*}
we have 
\begin{equation*}
\rho_{N_{o}} \leq \lambda \rho_0 .
\end{equation*}
The above inequality also implies that $\rho_{N_{o}} < \bar{\rho}_2$. Thus, the uniform boundedness in Claim \ref{Claim on the observability along the estimate trajectory} holds and the above process can be repeated to derive the following:
\begin{equation*}
\rho_{kN_o} \leq \lambda \rho_{(k-1)N_o} \leq \cdots \leq \lambda^k \rho_0 
\end{equation*}
which is similar to the result given in \cite{Shamma_Tu_1997}. This clearly establishes the asymptotic convergence property, i.e., $\lim_{k \rightarrow \infty} \rho_k = 0$.

Finally, we note that the above analyses also imply boundedness of the correction ellipsoid shape matrices. To this end, we note that $\bm{\nu} \equiv \bm{E}_{k|k} \bm{z}_{k|k}$ and $\rho_k \equiv \gamma_{k|k}$ where $\bm{E}_{k|k}$, $\bm{z}_{k|k}$, and $\gamma_{k|k}$ are as in Section \ref{Preliminaries and Problem Formulation}. Now, consider the case of nondivergence. Since $\rho_k$ is uniformly bounded, so is $\gamma_{k|k}$. This implies that the correction ellipsoid shape matrices remain uniformly bounded. Next, consider the asymptotic convergence case. For this, $\lim_{k \rightarrow \infty} \rho_k = 0$ means $\lim_{k \rightarrow \infty} \bm{z}_{k|k} = \bm{0}_n$. Then, due to the nature of set-membership filtering technique (i.e., at every time step, the correction ellipsoid is synthesized by solving a convex optimization problem that guarantees to contain the true state with the corrected state estimate at the corresponding center), we again have $\gamma_{k|k}$ bounded. A similar set of arguments can be made for the prediction ellipsoid shape matrices as well. This completes our discussion on the theoretical properties of the SDC-SMF for system \eqref{SDC form of the simplified nonlinear system}. 
\section{Simulation Example}  \label{Simulation Example}
A simulation example is provided in this section to illustrate the effectiveness of the proposed approach. All the simulations are carried out on a laptop computer with 8.00 GB RAM and 1.60-1.80 GHz Intel(R) Core(TM) i5-8250U processor running MATLAB R2019b. The SDPs in 
\eqref{The complete problem statement-1} and \eqref{The complete problem statement-2} are solved utilizing `YALMIP' \cite{Lofberg_2004} with the `SDPT3' solver in the MATLAB framework. 

Let us consider the Van der Pol equation in \cite{Shamma_Tu_1997} and express the discrete-time system as
\begin{equation*}
\begin{split}
\bm{x}_{k+1} &= \begin{bmatrix}
x_{1_{k}} + \Delta t x_{2_{k}} \\
x_{2_{k}} + \Delta t (-9 x_{1_{k}} + \mu (1-x_{1_{k}}^2) x_{2_{k}})
\end{bmatrix} + \begin{bmatrix}
0 \\ w_k
\end{bmatrix}, \\
&= \bm{f}_d (\bm{x}_k) + \bm{w}_k, \\
y_k &= x_{1_{k}} + v_k = \bm{h}_d (\bm{x}_k) + v_k
\end{split}
\end{equation*}
where $\bm{x}_{(\cdot)} = [x_{1_{(\cdot)}} \quad x_{2_{(\cdot)}}]^\text{T}$ and $\Delta t$ is the discretization time step. Clearly, the functions in the above system satisfy Assumption \ref{Assumptions on the nonlinear functions}. Then, utilizing \eqref{SDC form calculation-2}, we have
\begin{displaymath}
\begin{split}
\bm{A}(\bm{x}_k) &= \begin{bmatrix}
1 & \Delta t \\
 - 9 \Delta t - \frac{2}{3} \mu \Delta t x_{1_{k}} x_{2_{k}}  & 1 + \mu \Delta t (1 - \frac{1}{3} x_{1_{k}}^2)
\end{bmatrix} \\
\bm{H} (\bm{x}_k) &= \begin{bmatrix}
1 & 0
\end{bmatrix}.
\end{split}            
\end{displaymath}  
With these SDC matrices, we have
\begin{equation}
\mathcal{O}_{k,k+1} = \begin{bmatrix}
1  &   0 \\
1  &  \Delta t
\end{bmatrix}
\end{equation}
which is full-rank for all $\Delta t \neq 0$. Thus, the rank condition in \eqref{Observability rank condition} is satisfied with $N_o = 2$. We take $\mu = 2$ for which the Van der Pol equation (nominal part) admits a unique and stable limit cycle, thus satisfying Assumption \ref{State dynamics compactness assumption}. Also, we set $\Delta t = 0.05$ seconds and use $N = 1000$ for calculating $r_{A_{k}}$. With the above SDC parameterizations, the matrices $\bm{K}_1$ and $\bm{K}_2$ are given by
\begin{displaymath}
\begin{split}
\bm{K}_1 &= \begin{bmatrix}
0 & 0 & 0 & 0
\end{bmatrix} \\
\bm{K}_2 &= \begin{bmatrix}
0 & 0 & 0 & 0 \\
- \frac{2}{3} \mu \Delta t \hat{x}_{2_{k|k}} & - \frac{2}{3} \mu \Delta t \hat{x}_{1_{k|k}}  & - \frac{2}{3} \mu \Delta t \hat{x}_{1_{k|k}} & 0
\end{bmatrix} .
\end{split}
\end{displaymath}
\begin{figure}[!hbt]
\centering
\includegraphics[width= 1\columnwidth]{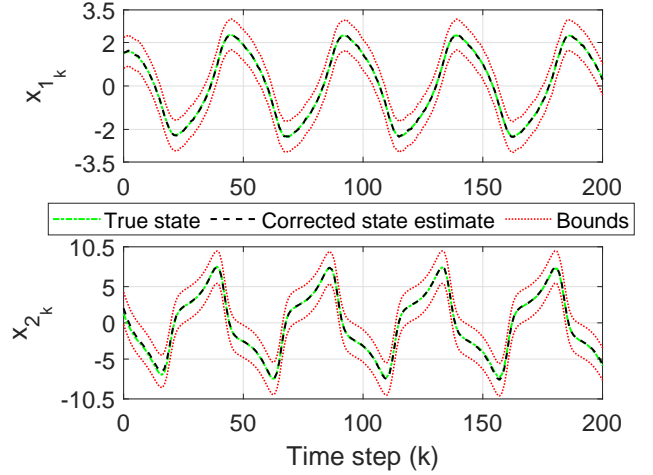}
\caption{Simulation results corresponding to the SDC-SMF.} 
\label{Example_1_1}
\end{figure}
\begin{figure}[!hbt]
\centering
\includegraphics[width= 1\columnwidth]{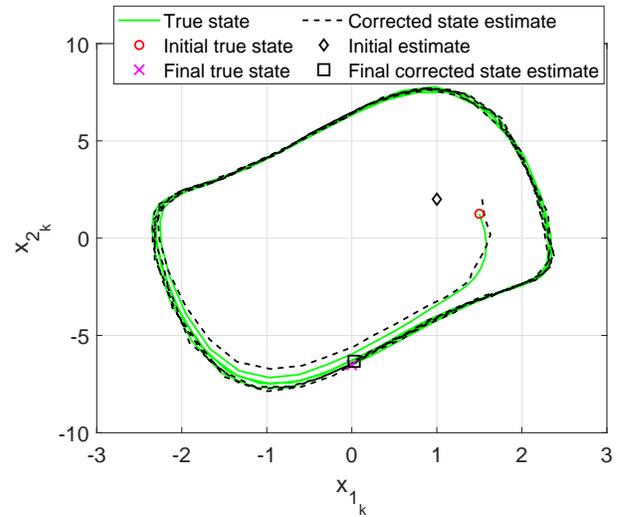}
\caption{True state and corrected state estimate trajectories in the phase plane.} 
\label{Example_1_2}
\end{figure}
In this example, the initial condition is given by $\bm{P}_0 = \bm{I}_2$, $\bm{x}_0 = [1.5 \quad 1.25]^{\text{T}}$, and $\hat{\bm{x}}_0 = [1 \quad 2]^{\text{T}}$. For Assumption \ref{Assumption 3 - initial conditions}, we can consider $\mathbb{D}_0 = \mathcal{E} (\hat{\bm{x}}_0, \bm{P}_0)$. In terms of Assumption \ref{Assumption 3 - process and measurement noise ellipsoids}, let us choose $q = r = 0.01$. Then, Assumption \ref{Assumption 3 - process and measurement noise ellipsoids} is satisfied with (i) $w_k$ and $v_k$ randomly varying (uniform distribution) between -0.05 and 0.05; (ii) $\bm{Q}_k = 0.01 \bm{I}_2$, $\bm{R}_k = 0.01$.
The true state components along with the corresponding corrected state estimates and bounds are shown in Fig. \ref{Example_1_1} as functions of time steps. Clearly, $x_{1_{k}}$, $x_{2_{k}}$ remain within the bounds for the entire time-horizon considered which mean that the true state is successfully contained in the correction ellipsoids. Fig. \ref{Example_1_2} depicts the true state trajectory and the corrected state estimate trajectory in the phase plane. Note that, at $k=0$, the correction step brings the corrected state estimate close to the initial true state. Also, it is obvious that the corrected state estimate trajectory converges to and remains in a neighborhood of the true state trajectory after a few recursions of the filter. 
\begin{table}[!hbt] 
\centering
\caption{Mean trace and estimation error comparisons over 200 time steps} \label{Table - Comparison}
\begin{tabular}{|c|c|c|} 
\hline
Item & SDC-SMF & Wang et al. \cite{Wang_et_al_2018} \\ \hline
Mean trace & 5.5007 & 6.2616 \\ \hline
MAE & 0.1142 & 0.1761  \\ \hline
MSE & 0.0277 &  0.0643 \\ \hline
\end{tabular}
\end{table}
\begin{figure}[!hbt]
\centering
\includegraphics[width= 1\columnwidth]{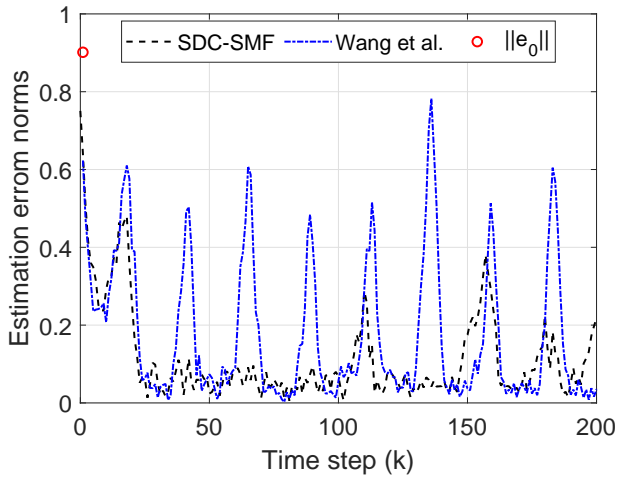}
\caption{Estimation error norms for the SDC-SMF and the SMF in \cite{Wang_et_al_2018} (Wang et al.).} 
\label{Example_1_Comparison-1}
\end{figure}

\begin{figure}[!hbt]
\centering
\includegraphics[width= 1\columnwidth]{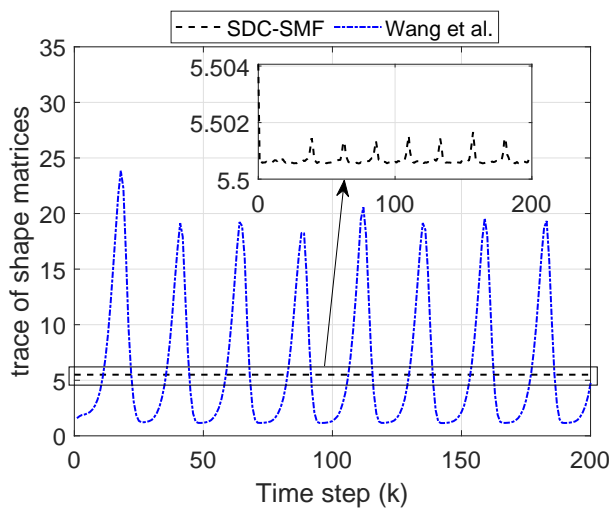}
\caption{Trace of correction ellipsoid shape matrices for the SDC-SMF and state estimation ellipsoid shape matrices for the SMF in \cite{Wang_et_al_2018} (Wang et al.). } 
\label{Example_1_Comparison-2}
\end{figure}
Next, for comparison, we implement the SMF in \cite{Wang_et_al_2018} for the above example with the remainder bounding ellipsoids synthesized using 50 constraints. Let us consider the estimation errors at the correction steps for the SDC-SMF and at the measurement update steps for the SMF in \cite{Wang_et_al_2018}. The comparison in these estimation error norms is shown in Fig. \ref{Example_1_Comparison-1} where $||\bm{e}_0|| = ||\bm{x}_0 - \hat{\bm{x}}_0||$ is the initial error norm and the comparison in trace of the corresponding ellipsoid shape matrices is shown in Fig. \ref{Example_1_Comparison-2}. The results in Figs. \ref{Example_1_Comparison-1}, \ref{Example_1_Comparison-2} demonstrate that the SDC-SMF outperforms the SMF in \cite{Wang_et_al_2018}. This is further illustrated in the results given in Table \ref{Table - Comparison} where MAE and MSE stand for mean absolute error and mean squared error, respectively. The SDC-SMF performs much better in terms of these two metrics, as shown in Table \ref{Table - Comparison}. Also, the mean trace value for the SDC-SMF correction ellipsoid shape matrices is smaller compared to that of the state estimation ellipsoid shape matrices for the SMF in \cite{Wang_et_al_2018}. In summary, the SDC-SMF results in lower estimation errors with lower error bounds for this example. 

\begin{figure}[!hbt]
\centering
\includegraphics[width= 1\columnwidth]{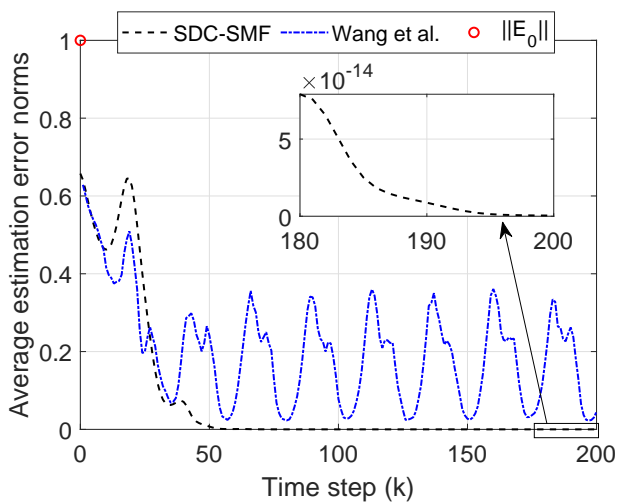}
\caption{Average estimation error norms with $w_k = v_k = 0$ and $n_r = 10$.} 
\label{Example_1_Comparison-4}
\end{figure}

Finally, to demonstrate that the SDC-SMF is asymptotically convergent if there are no process and measurement noises (cf., Section \ref{Theoretical properties of SDC-SMF}), we implement the SDC-SMF for the above example with the initial state randomly chosen from the boundary of the initial ellipsoid $\mathcal{E} (\hat{\bm{x}}_0, \bm{P}_0)$ and with $w_k = v_k = 0$. We repeat this process $n_r$ times. The same is done for the SMF in \cite{Wang_et_al_2018} as well.
The average estimation error norms of these runs with the random initializations are shown in Fig. \ref{Example_1_Comparison-4} where $\bm{E}_0$ is the Cholesky factorization of $\bm{P}_0$. Note that the upper bound of the initial error norm for the random initializations is $||\bm{E}_0||$, which is shown in Fig. \ref{Example_1_Comparison-4}. The results in Fig. \ref{Example_1_Comparison-4} show that the SDC-SMF is asymptotically convergent with the estimation error tending to zero. However, the SMF in \cite{Wang_et_al_2018} does not exhibit this property, as shown in Fig. \ref{Example_1_Comparison-4}. 
\section{Conclusion} \label{Conclusion}
A recursive set-membership filtering algorithm for discrete-time nonlinear dynamical systems subject to unknown but bounded process and measurement noise has been derived utilizing the state dependent coefficient (SDC) parameterization. At each time step, the filtering problem has been transformed into two semi-definite programs (SDPs) using the S-procedure and Schur complement. Optimal (minimum trace) ellipsoids have been constructed that contain the true state of the system at the correction and prediction steps. Finally, a simulation example is provided which demonstrates that the proposed filter performs better compared to an existing set-membership filter for discrete-time nonlinear systems. Our future research would involve assessing theoretical properties of the SDC-SMF for systems with control inputs acting through a possibly non-square state-dependent matrix.
\section*{Acknowledgment}
This research was supported by the Office of Naval Research under Grant No. N00014-18-1-2215. The authors would like to thank the anonymous reviewers for their suggestions which lead to improved quality and presentation of the technical note.

\bibliography{Bibliography}
\bibliographystyle{IEEEtran}

\end{document}